\documentclass[apj]{emulateapj}
\usepackage{multirow}
\usepackage{graphics}

\shorttitle{Demographics of SDSS galaxies}
\shortauthors{Oh et al.}

\begin{document}

\title{Demographics of Sloan Digital Sky Survey Galaxies Along the Hubble Sequence}

\author{Kyuseok Oh\altaffilmark{1}, Hyunseop Choi\altaffilmark{1}, Hong-Geun Kim\altaffilmark{1}, Jun-Sung Moon\altaffilmark{1} and Sukyoung K. Yi\altaffilmark{1,2}}
\altaffiltext{1}{Department of Astronomy, Yonsei University, Seoul 120-749, Korea; yi@yonsei.ac.kr} 
\altaffiltext{2}{Yonsei University Observatory, Yonsei University, Seoul 120-749, Korea} 
\altaffiltext{1}{yi@yonsei.ac.kr}

\def\OI{[\mbox{O\,{\sc i}}]~$\lambda 6300$}
\def\OIII{[\mbox{O\,{\sc iii}}]~$\lambda 5007$}
\def\OIIIs{[\mbox{O\,{\sc iii}}]~$\lambda 4363$}
\def\OIIIab{[\mbox{O\,{\sc iii}}]$\lambda\lambda 4959,5007$}
\def\SIIab{[\mbox{S\,{\sc ii}}]~$\lambda\lambda 6717,6731$}
\def\SII{[\mbox{S\,{\sc ii}}]~$\lambda \lambda 6717,6731$}
\def\NII{[\mbox{N\,{\sc ii}}]~$\lambda 6584$}
\def\NIIb{[\mbox{N\,{\sc ii}}]~$\lambda 6584$}
\def\NIIa{[\mbox{N\,{\sc ii}}]~$\lambda 6548$}
\def\NI{[\mbox{N\,{\sc i}}]~$\lambda \lambda 5198,5200$}

\def\OIIa{[\mbox{O{\sc ii}}]~$\lambda 3726$}
\def\OIIb{[\mbox{O{\sc ii}}]~$\lambda 3729$}
\def\NeIIIa{[\mbox{Ne{\sc iii}}]~$\lambda 3869$}
\def\NeIIIb{[\mbox{Ne{\sc iii}}]~$\lambda 3967$}
\def\OIIIa{[\mbox{O{\sc iii}}]~$\lambda 4959$}
\def\OIIIb{[\mbox{O{\sc iii}}]~$\lambda 5007$}
\def\HeII{{He{\sc ii}}~$\lambda 4686$}
\def\ArIVa{[\mbox{Ar{\sc iv}}]~$\lambda 4711$}
\def\ArIVb{[\mbox{Ar{\sc iv}}]~$\lambda 4740$}
\def\NIa{[\mbox{N{\sc i}}]~$\lambda 5198$}
\def\NIb{[\mbox{N{\sc i}}]~$\lambda 5200$}
\def\HeI{{He{\sc i}}~$\lambda 5876$}
\def\OI{[\mbox{O{\sc i}}]~$\lambda 6300$}
\def\OIb{[\mbox{O{\sc i}}]~$\lambda 6364$}
\def\SIIa{[\mbox{S{\sc ii}}]~$\lambda 6716$}
\def\SIIb{[\mbox{S{\sc ii}}]~$\lambda 6731$}
\def\ArIII{[\mbox{Ar{\sc iii}}]~$\lambda 7136$}

\def\Ha{{H$\alpha$}}
\def\Hb{{H$\beta$}}

\def\NIIHa{[\mbox{N\,{\sc ii}}]/H$\alpha$}
\def\SIIHa{[\mbox{S\,{\sc ii}}]/H$\alpha$}
\def\OIHa{[\mbox{O\,{\sc i}}]/H$\alpha$}
\def\OIIIHb{[\mbox{O\,{\sc iii}}]/H$\beta$}

\def\Ebmv{E($B-V$)}
\def\LOIII{$L[\mbox{O\,{\sc iii}}]$}
\def\Ledd{${L/L_{\rm Edd}}$}
\def\LOIIIs4{$L[\mbox{O\,{\sc iii}}]$/$\sigma^4$}
\def\LOIIIMbh{$L[\mbox{O\,{\sc iii}}]$/$M_{\rm BH}$}
\def\Mbh{$M_{\rm BH}$}
\def\Msigma{$M_{\rm BH} - \sigma$}
\def\Ms{$M_{\rm *}$}
\def\Msun{$M_{\odot}$}
\def\Msunyr{$M_{\odot}yr^{-1}$}

\def\ergs{$~\rm ergs^{-1}$}
\def\kms{${\rm km}~{\rm s}^{-1}$}
\newcommand{\cms}{\mbox{${\rm cm\;s^{-1}}$}}
\newcommand{\pccm}{\mbox{${\rm cm^{-3}}$}}
\newcommand{\sauron}{{\texttt {SAURON}}}
\newcommand{\oasis}{{\texttt {OASIS}}}
\newcommand{\HST}{{\it HST\/}}

\newcommand{\Vg}{$V_{\rm gas}$}
\newcommand{\Sg}{$\sigma_{\rm gas}$}
\newcommand{\eg}{e.g.,}
\newcommand{\ie}{i.e.,}

\newcommand{\gandalf}{{\texttt {gandalf}}}
\newcommand{\fracDeV}{{\texttt {FracDeV}}} 
\newcommand{\ppxf}{{\texttt {pPXF}}}

\begin{abstract}

We present the statistical properties of a volume-limited sample of 7,429 nearby ($z = 0.033$ -- 0.044) galaxies from the Sloan Digital Sky Survey Data Release 7. Our database includes morphology distribution as well as the structural and spectroscopic properties of each morphology type based on the recent re-measurements of spectral line strengths by Oh and collaborators (2011). Our database does not include galaxies that are apparently smaller and flatter, because morphology classification of them turned out to be difficult. Our statistics confirmed the up-to-date knowledge of galaxy populations, e.g., correlations between morphology and line strengths as well as the derived ages, etc. We hope that this database will be useful as a reference.
\end{abstract}	


\section{Introduction}

Three quarters of a century after the introduction of the Hubble Sequence \citep{hub26, hub36}, galaxy morphology is still a mystery and the subject of numerous studies. Morphology classification is often the first step in many galaxy-related investigations. Regarding the implications of the Hubble Sequence, galaxy morphology is considered variable, as it was in the original impression in Hubble's scheme. However, unlike in the original impression, morphology transformation is currently considered possible in many different ways.

The original, simple, two-pronged classification has since been modified and expanded to suit more galaxies discovered \citep{dev59, san61}. Based on the HubbleÕs tuning fork scheme, \citet{van60a, van60b, van60c} suggested his classification, called the DDO system, considering galaxy luminosity and the presence of spiral arms. In this scheme, luminosity class I are supergiant galaxies with long and well-developed spiral arms. Luminosity class III are giant galaxies with patchy arms, and luminosity class V are dwarf galaxies of very low surface brightness showing only a hint of spiral structure. Later, \citet{van76} further elaborated his scheme by adding a parallel sequence of anemic spirals and gas-free lenticular galaxies according to the amount of gas in the disk. On the other hand, \citet{kor79} utilized additional shape parameters such as lens (lenticular galaxies) and ring structure as secondary classification criteria, based on 121 galaxies listed in the \textit{Second Reference Catalogue of Bright Galaxies} \citep{dev76}. In addition, \citet{dev74} presented various correlations between morphology and other galaxy properties, such as luminosity, ellipticity, mass, density, spectral energy distribution, and colors. It is also worth noting that \citet{dre80} conducted a study that was based on a dramatically increased number of samples. The study, which was based on roughly 6,000 bright galaxies from 55 clusters, unveiled a tight correlation between density and morphology. Later, \citet{rob94} showed the morphological dependence of fundamental galaxy properties according to the Hubble types using the Third Reference Catalogue of Bright Galaxies (RC3, \citealt{dev91}) and the Arecibo General Catalogue. Based on these previous works, \citet{kor96} proposed a revised version of HubbleÕs tuning fork, particularly for early-type galaxies, using the velocity anisotropy that can be deduced from the isophotal shapes of galaxies. Meanwhile, \citet{mor58, mor59, mor62} developed a fundamentally different morphology classification system based on the central concentration of light. This one-dimensional classification scheme, often referred to as the Yerkes system, arranged galaxies in its \emph{a-f-g-k} sequence from the weakest central concentration of light (\emph{a}) to the strongest (\emph{k}). The connection between morphology and central concentration of light was later confirmed by numerous studies including \citet{abr94}, \citet{ber00}, \citet{shi01} and \citet{con03}.

\begin{figure}
\centering
\includegraphics[width=0.5\textwidth]{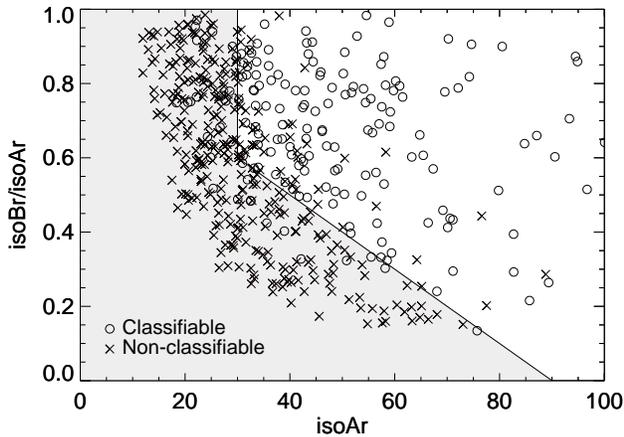}
\caption{The apparent shape criteria in our sample selection based on 500 random galaxies. 
The $\rm IsoA_r$ and $\rm IsoB_r$ indicate the semi-major and semi-minor axes of each apparent galaxy image. 
Circles show the galaxies we felt confident about the morphology classification for, whereas crosses show the non-classifiable galaxies. We concluded that using the SDSS images, small (in the semi-major axis) or low $\rm IsoB_r$/$\rm IsoA_r$ galaxies were difficult to classify. Therefore, we set up an initial sample by selecting only the galaxies whose apparent shapes fell in the white area of this diagram to make our exercise feasible. These criteria removed approximately 70\% of the galaxies.}
\label{sample}
\end{figure}

With the advent of mega-scale surveys that provide pipeline-measured properties, further considerations of galaxy classification became possible. Studies on the Hubble Space Telescope Medium Deep Survey \citep{abr96} and the 2 degree Field Galaxy Redshift Survey \citep{col01} were good examples, and now it is more readily possible  through the Sloan Digital Sky Survey (SDSS, \citealt{yor00}). For example, \citet{abr03} applied the asymmetry-related Gini coefficient as a new tool for galaxy morphology classification. From a different direction of effort, asymmetry and clumpiness in the galaxy image have been used to classify galaxies (e.g., \citealt{abr94, abr96, con03, lot04}). 

Another great advantage of using mega-scale surveys is that we can achieve results that are statistically more significant. The most recent SDSS database contains roughly a million galaxies, which is orders of magnitude larger than older databases. Such large modern surveys provide galaxy properties (e.g., concentration index and \fracDeV\ as discussed in Section 3) measured in uniform ways and, therefore, make many interesting investigations possible \citep{ber03, sch07, lin08, suh10, tem11, gon11}. In regards to the size of the database for morphology classification, it should also be noted that \citet{fuk07}, \citet{nai10} and the EFIGI team \citep{bai11, de11} performed detailed morphology classification using the SDSS DR3, DR4 and RC3, respectively (see Section 3.2 for details).

In this study, we aimed to derive the representative properties of each Hubble class with statistical significance using the SDSS Data Release 7 (DR7, \citealt{aba09}). First of all, unlike most of the previous studies including those mentioned above, we set up an unbiased {\em volume-limited sample} with a restricted redshift range. We morphologically classified the galaxies into the Hubble Sequence and inspected their mean properties such as colors, magnitudes, stellar mass, velocity dispersion, local number density, and their stellar absorption- and nebular emission-line strengths. For spectroscopic information, we adopted the measurements of the OSSY catalogue \citep{oh11}. The OSSY database provides new line strength measurements based on advanced spectral fits and quality assessments. The strengths of our investigation were twofold: our database is a volume-limited sample, and, second, our spectroscopic statistics are based on the OSSY catalogue.

This paper is organized as follows. In Section 2, we describe the sample selection criteria. In Section 3, we explain how we classified the galaxy morphology. After introducing the galaxy concentration index and the fraction of de Vaucouleurs profile as a sanity check, the color-magnitude relationship, distributions of the stellar mass, and velocity dispersion are presented. The stellar absorption-line and nebular emission-line features are then described in Sections 4 and 5, respectively. We discuss density against galaxy morphology in Section 6 and give a summary of our work in Section 7.

\begin{figure*}
\centering
\includegraphics[width=1\textwidth]{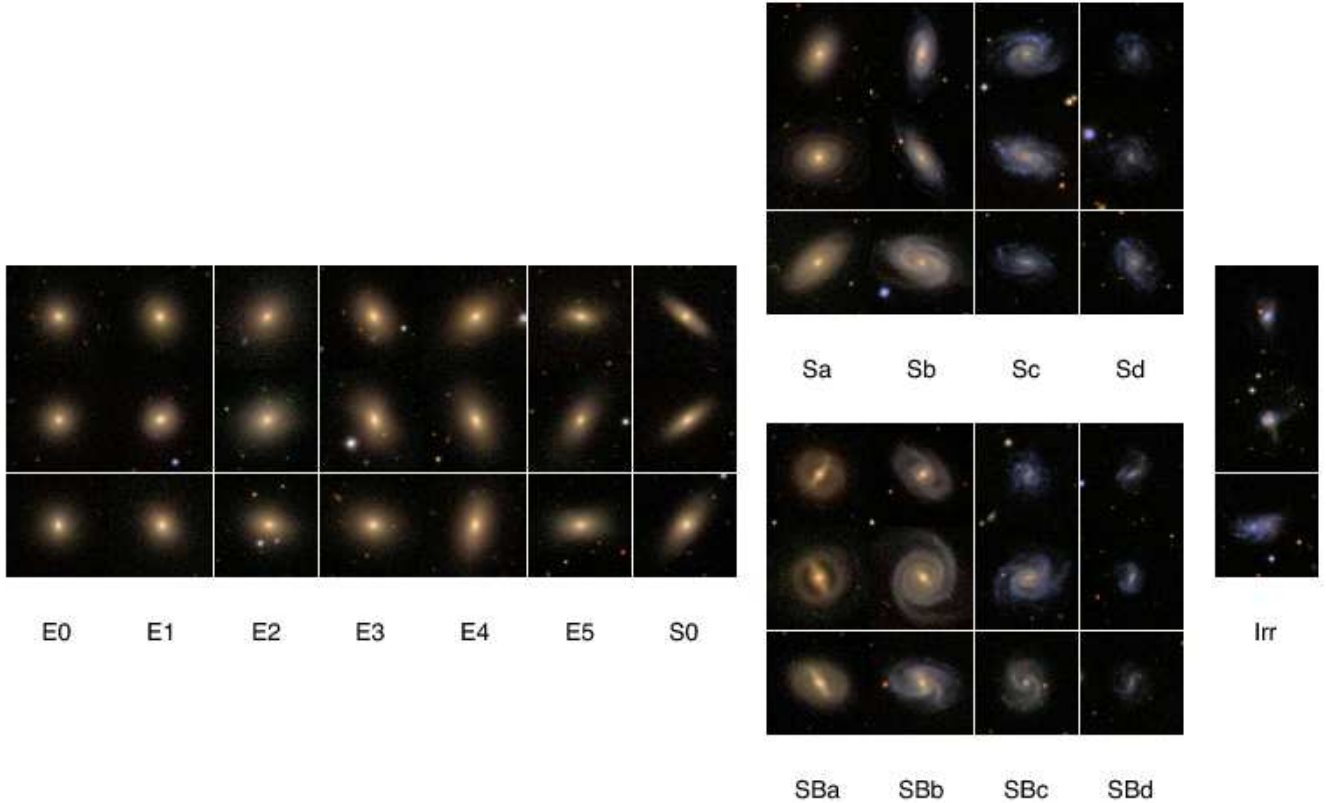}
\caption{Samples of the SDSS DR7 composite images of morphologically classified galaxies in the Hubble Sequence. 
Each class shows 3 sample galaxies.}
\label{morphology}
\end{figure*}

\section{Sample Selection}
\label{sec:sample}

We extracted galaxies at the redshift range, $0.033 < z < 0.044$, from the SDSS DR7 database. This method yielded 32,868 galaxies. We felt that this number was large enough to serve as a starting pool of sample galaxies. Such a tight redshift cut helped us to avoid the evolution effect. Our volume-limited sample had a magnitude limit of $M_{\rm r} \leq -18.65$, assuming the WMAP 7-yr $\Lambda$CDM cosmology \citep{kom11}, which is considerably fainter than most of the previous studies mentioned in Section 1.

We first performed visual morphology classification on a subsample of 500 random galaxies and found that classification is practically impossible when a galaxy is too small or having a low $IsoB_r$/$IsoA_r$ in appearance. Most of the non-classifiable galaxies (68$\%$) were located in the gray area in Fig.~\ref{sample}. While it is admittedly subjective, we determined the demarcation lines based on the semi-major ($\rm IsoA_r$) and semi-minor ($\rm IsoB_r$) axes of the galaxy images, as shown in Fig.~\ref{sample}. In summary, we excluded the galaxies of size $\rm IsoA_r < 30$" or ${\rm IsoB_r} / {\rm IsoA_r} < -0.01 \, {\rm IsoA_r} + 0.90$, after which 10,233 galaxies remained (only approximately 30\% of the initial sample). Eliminating all of the apparently small and low $IsoB_r$/$IsoA_r$ galaxies, our sampling inevitably suffered from a bias against such galaxies. As a result, our sample does not properly represent intrinsically low $IsoB_r$/$IsoA_r$, early-type galaxies, such as E5 through E7. This is an important caveat of our investigation. After extensive visual inspection of the 10,233 galaxies, we felt reasonably confident in our morphological classification of only 7,433 galaxies. Our selection scheme removes 84.2 \% our ``non-classifiable galaxies'' and retains 81.5 \% of ``classifiable galaxies''. It is worth noting that none of the morphology type preferentially occupy the specific zone in Fig.~\ref{sample}.


\section{Basic Properties of Galaxies}
\label{sec:basic}

\subsection{Morphology}
\label{sec:morphology}

Morphology classification was conducted using the SDSS \textit{gri} composite color images with two different image scales (72 $"\times72$", 216 $"\times216$"). It should be noted that using a fixed contrast ratio for the galaxy morphology classification is difficult to distinguish elliptical galaxies from S0. Color itself was not used as a classification criterion.

The existence of disk feature played the most important role for distinguishing disk galaxy from elliptical galaxy. Depending on the visibility of arm structures, disk galaxies were divided into lenticular or spiral galaxy. Following the guidelines from \citet{hub26}, from Sa to Sd, the spiral galaxy's subclass was {\em visually} determined primarily by the tightness of the galaxy's arms and the relative bulge dominance. Identification of barred spirals was also done visually and, therefore, was subject to a substantial uncertainty. We did not follow measure bar strength or length but just confirmed the presence of bar through image inspection. It is interesting to note that \citet{oh12} claimed that visual identification is still more effective than automated techniques, at least for SDSS images. SBd galaxies were difficult to identify, because they are generally smaller and fainter than earlier type galaxies and also because their bulges are small, while bar size usually correlates with bulge size \citep{oh12}. Consequently, only 18 SBd galaxies were found and the statistics for the SBd might not be reliable.

Elliptical galaxy classification was done in two steps. First, we identified the galaxies as elliptical when it shows no disk feature and relatively continuous light profile without any clumpy structures. Second, we used $\rm IsoA_r$ and $\rm IsoB_r$ from SDSS in order to determine their apparent ellipticity. The $\rm IsoA_r$ and $\rm IsoB_r$ measurements were from the projected shapes of galaxies and, therefore, may not reflect their true shapes \citep{kim07}. Due to our sampling strategy described in Section 2, only a small number of galaxies were labeled as E5 (39), E6 (4), or E7 (0). We found it meaningless to derive mean properties for a sample with 4 galaxies and, therefore, removed the four E6 galaxies from our sample. This brought the number in our final sample down to 7,429.

Elliptical and S0 (and sometimes even Sa) galaxies appeared very similar to each other \citep{van09}, which made it difficult to distinguish them from each other. For much of our analysis, we combined elliptical and S0 galaxies into ``early-types'' for this reason. Fig.~\ref{morphology} shows sample galaxy images along the Hubble Sequence.

Irregular galaxies were classified by the feature of unclear nucleus structure and the degree of broken symmetry. Some of the apparently-small galaxies do not show clear features such as symmetrical spiral arms or central bulge and hence were classified as irregulars.

``Unknown'' type galaxies are different from the non-classifiable ones. After filtering out non-classifiable galaxies through the method described in Section 2 (Fig.~\ref{sample}), 2,894 out of 10,233 galaxies were defined as unknown by visual classification. As described in the following section, most of our unknown-type galaxies are classified as S0 by other previous investigators. The rest consists of relatively more inclined galaxies and hence more difficult to classify and tidally-disturbed galaxies making it difficult to determine their detailed sub-types.

\begin{figure}
\centering
\includegraphics[width=0.5\textwidth]{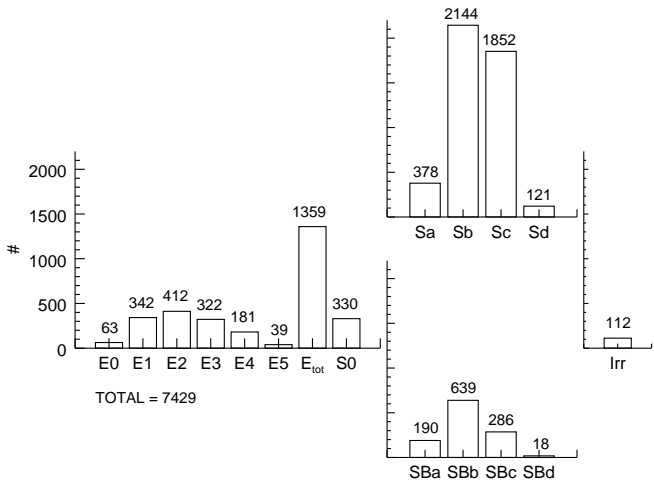}
\caption{Morphology classification result in the format of the Hubble Sequence. 
The number of galaxies in each classification is written above each bar. 
Detailed statistics are given in Tab.~\ref{tab:tabletwo}, Tab.~\ref{tab:tablethree}, Tab.~\ref{tab:tablefour} and Tab.~\ref{tab:tablefive}.}
\label{number}
\end{figure}

Fig.~\ref{number} shows the number distribution of galaxies along the Hubble Sequence. In the final sample, 22.7\% (1,689), 75.8\% (5,628), and 1.5\% (112) of the galaxies were classified as early-type, late-type, and irregular, respectively. These values were more uncertain for later types of galaxies for several reasons including the caveat in our sampling strategy. Yet, we still believe that these galaxies are useful for providing rough ideas of the true morphology mix in the local universe. The disk dominance was primarily due to the fact that our luminosity cut was fairly faint, while disk galaxies tend to be fainter on average than early-type galaxies.

\begin{figure}
\centering
\includegraphics[width=0.5\textwidth]{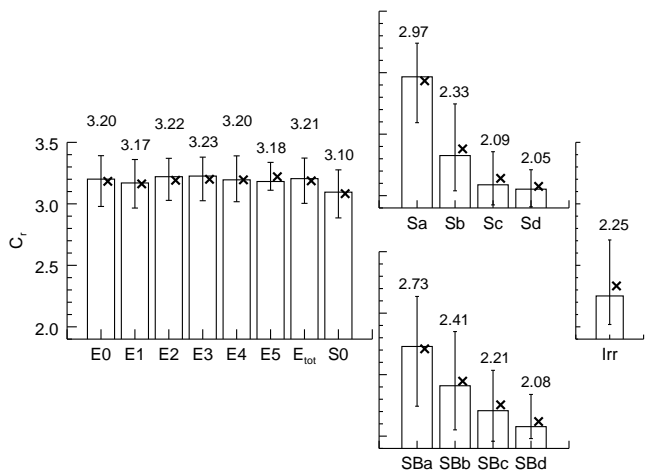}
\caption{Concentration index. The bars and the associated error bars represent 
the median values with the standard deviations, while the x symbols show the mean.}
\label{cr}
\end{figure}

\begin{figure}
\centering
\includegraphics[width=0.5\textwidth]{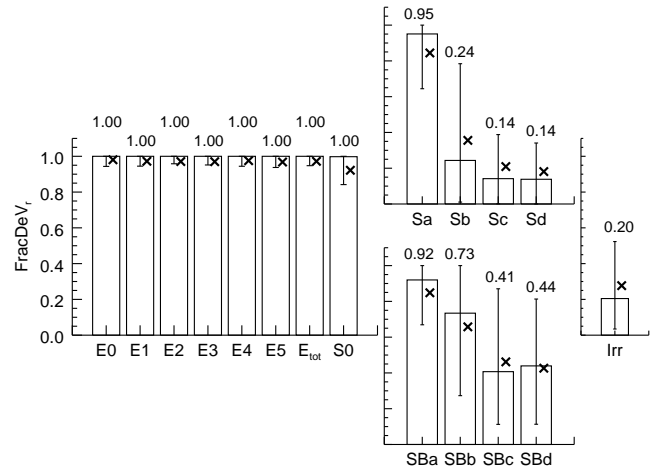}
\caption{Fraction of de Vaucouleurs profile (\fracDeV) in the r-band brightness profile of the galaxy. 
The format is the same as that of Fig. 4.}
\label{fracdev}
\end{figure}

\begin{figure*}
\centering
\includegraphics[width=1\textwidth]{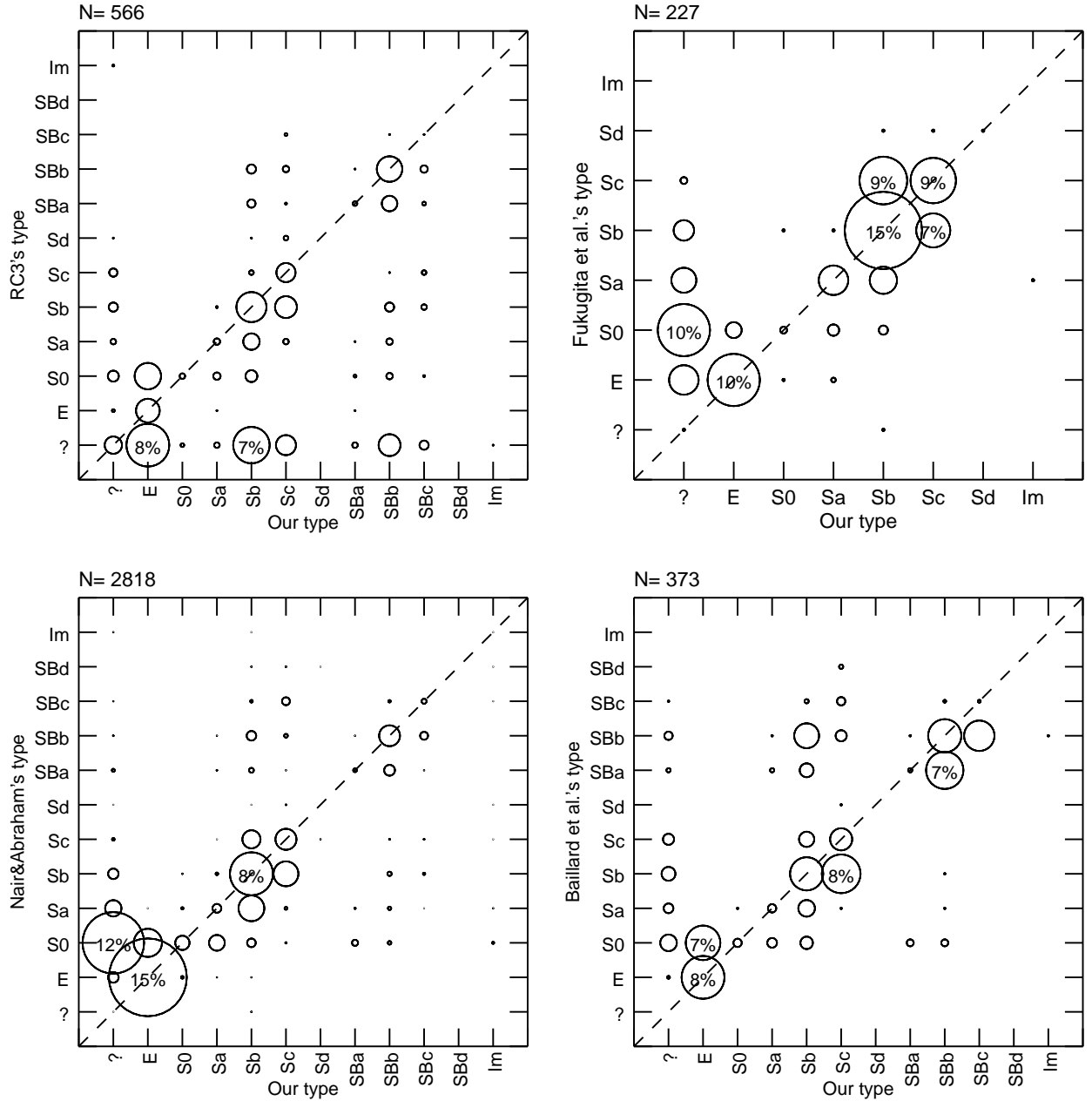}
\caption{Comparison with previous works. 
Size of circle represents the fraction. 
Only the fractions larger than 6.5 per cent are noted with numerics. 
Number of objects in common between catalogues is listed on top-left at each panel.
Question mark stands for ``unknown'' class.}
\label{comp}
\end{figure*}

We checked the reliability of our morphology classification by inspecting the concentration index and the \fracDeV\ provided by SDSS. The concentration index has been widely used in galaxy morphology classification \citep{doi93, abr94, abr96, shi01, con03}. In this study, we used the concentration index that is defined as the ratio between PetroR90 and PetroR50. In other words, $C_{\rm r} \equiv \rm PetroR90/PetroR50$, from $r$-band petrosian magnitudes. A clear trend in the concentration index based on morphology is visible in Fig.~\ref{cr}. We should point out that in this figure, and some of the other following figures, we adopted the same format resembling the Hubble Sequence. It should be noted that we did not include E6 or E7 for the reasons we discussed in Sections 2 and 3. Since there is virtually no notable difference among the different subclasses of elliptical galaxies, we introduced a bar called ``$\rm E_{tot}$'' combining all of elliptical galaxies from E0 through E5 and we placed it between the ``E5'' and ``S0'' bars. Each bar shows the median with the standard deviation and the mean (x symbols) for reference.

Our elliptical galaxies had a concentration index higher than 3.0 with minimal dispersion, indicating a centrally concentrated light distribution. We did not see any notable distinction between the different subclasses in the elliptical category. Spiral galaxies (in the median values) showed a strong trend from Sa to Sd, which was consistent with the anticipated results from the increasing trend of the disk contribution from Sa to Sd. However, the dispersions were very large and, therefore the statistical significance may be low.

We performed a similar exercise using \fracDeV\ as well (Fig.~\ref{fracdev}). The SDSS pipeline parameter gives the fraction of light that is attributed to the de Vaucouleurs profile as opposed to the exponential disk profile. A pure bulge will have a value of 1, while a pure disk will have a value of 0. Our early-type galaxy classifications (E0 -- E5) had medians of 1.0, while spiral galaxies showed smaller values. The general trends were the same as those found in the concentration index test shown in Fig.~\ref{cr}.

Both of the tests performed based on the concentration index and \fracDeV\ indirectly indicated that our morphology classification was sensibly performed.

\subsection{Comparison with previous works}

We show in Fig.~\ref{comp} the comparison of morphology classification between select previous catalogues and ours. The top-left panel shows one-to-one correspondence relation comparing classifications for the 566 of our objects that are in the RC3. Size of circle  denotes the fraction in each class. For the accurate cross-matching process, we used the improved coordinates of HyperLeda database \citep{pat03}. By and large, our classifications are consistent with those of the RC3 as long as they are classified by both exercises. We are generally more conservative by putting more galaxies into the ``unknown'' category. Symmetrical minor discrepancies in classification are shown for (barred) spiral galaxies. This trend is consistently shown in other comparisons using the catalogues established by \citet{fuk07}, \citet{nai10}, and \citet{bai11}. Even though there are only 227 galaxies in common between \citet{fuk07} and our catalogue, correlation between the two catalogues is good. Based on the SDSS DR4 spectroscopic release, \citet{nai10} presented a morphological catalogue for 14,034 galaxies in the redshift range $0.01 < z < 0.1$ with apparent limit-magnitude of $g < 16$ mag. As we can expect, the SDSS based catalogue shows the largest overlap with our catalogue. The catalogue presented by \citet{bai11}, the EFIGI team, is a sub-sample of the RC3 and it naturally leads to a smaller number of cross-matched objects compared to the RC3. They classified 4,458 nearby galaxies at $0 < z \lesssim 0.05$ extracted from the Principal Galaxy Catalogue with supplemental information from the SDSS Data Release 4 (DR4). \citet{bai11} quantified bar length parameter with five types (0, 0.25, 0.5, 0.75 and 1). Bottom-right panel of Fig.~\ref{comp} was generated regarding 0.25 bar length as the representative barred galaxies. We also confirm that the level of agreement between these independent classifications is very good when we used concentration index as a reference parameter.

\subsection{Color-Magnitude Diagram}

\begin{figure}
\centering
\includegraphics[width=0.5\textwidth]{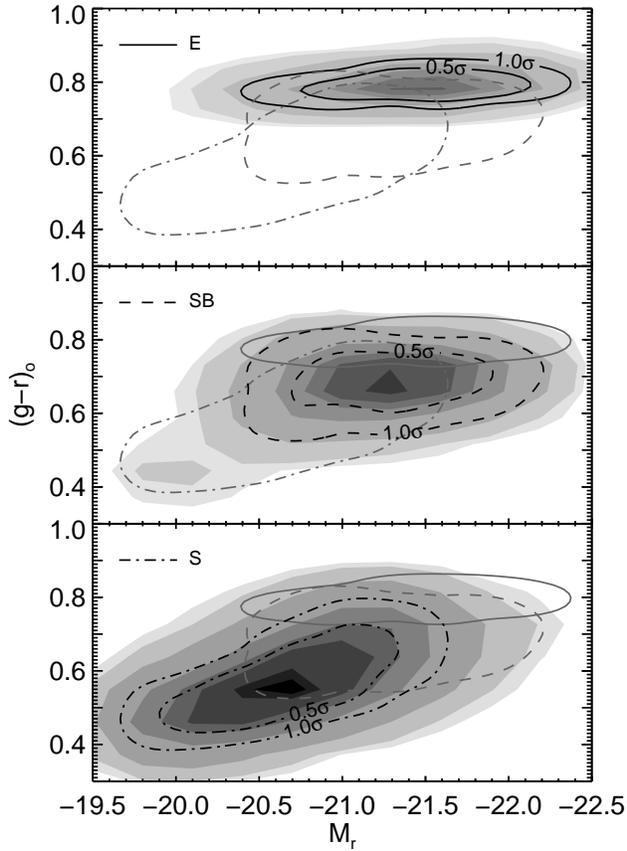}
\caption{Color-magnitude diagrams. 
The dark contours in each panel represent elliptical (solid lines, top), barred spiral (dashed lines, middle) and unbarred spiral galaxies (dot-dashed lines, bottom). 
Grey contours show the 1$\sigma$ contours of the other types.}
\label{cmr}
\end{figure}

Our early- and late-type galaxies populated distinctively different regions in the color-magnitude diagram, as shown in Fig.~\ref{cmr}. Early-type (E0 -- E5) galaxies exhibited a tight color-magnitude relationship as has been found in numerous studies in the past \citep{san78, bow92, dri06}. Late-type galaxies are fainter and bluer on average than early-type galaxies, forming the often-called Òblue cloudÓ. The bimodal separation is clear \citep{str01, bla03, hog03}. However, the distribution of barred spirals is less distinctive. A bar is found more often in earlier-type spiral galaxies (i.e. Sa, Sb) than in later-type spiral galaxies (Sc and Sd) in our volume-limited sample. Our classification is in good qualitative agreement with those of \citet{mas11} and \citet{oh12} who also used volume-limited samples \footnote{There have been reports that are inconsistent with our result. For example, \citet{bar08} showed that bluer galaxies have higher bar fractions. \citet{nai10} found two peaks in bar fraction at low and high masses. However, our sample shows lower bar fractions at bluer color and lower mass regimes. First of all, this discrepancy is partly due to the fact that detailed morphologies of low mass and blue galaxies are difficult to determine. Our barred spiral galaxies are well matched with ``strong'' and ``intermediate'' sub-classes of \citet{nai10}. On the other hand, we tend not to classify their ``weak'' galaxies as barred galaxies.  In addition, volume-limitation significantly affects on the statistics of low mass galaxies. If we apply the same volume limitation removing fainter galaxies from the sample, most studies agree with each other in the trends in bar fraction.}.

\subsection{Stellar Mass}

\begin{figure}
\centering
\includegraphics[width=0.5\textwidth]{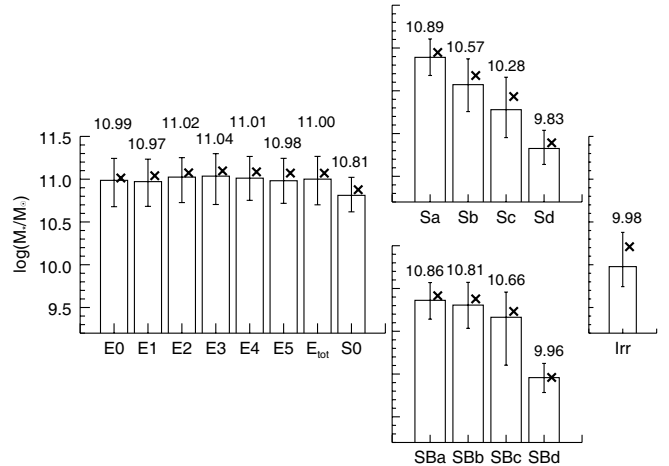}
\caption{Stellar mass. The format is the same as that of Fig.4.}
\label{mstar}
\end{figure}

\begin{figure}
\centering
\includegraphics[width=0.5\textwidth]{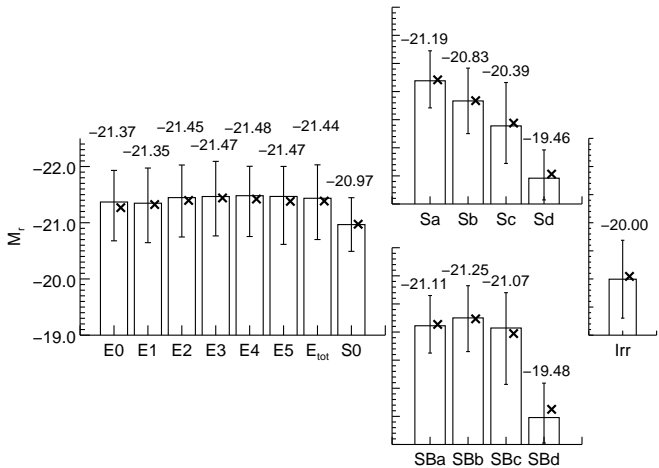}
\caption{Absolute magnitude. The format is the same as that of Fig.4.}
\label{absmag}
\end{figure}

We estimated the stellar mass of our galaxies using k-corrected colors and magnitudes following the formulae developed by Bell $\&$ de Jong (2001) and presented in Fig.~\ref{mstar}. The median of the stellar mass was higher for early-type galaxies, as expected. The median masses of early- (E0 through E5) and late- (S(B)b through S(B)d in this case) type galaxies were $10^{11.0}$ and  $10^{10.3}M_{\odot}$, respectively. We did not see much of a trend in stellar mass along the Hubble Sequence with the early-type galaxies. This result may imply that there is little type dependence of stellar mass among elliptical galaxies. However, this result could also be attributed to the fact that the apparent shape of elliptical galaxies does not rigorously reflect the true shape. The type dependence of stellar mass is clearer in the late-type galaxies. Again, this is probably a result of the fact that sub-classification is more straightforward for late-type galaxies. Fig.~\ref{absmag} shows the HubbleÕs tuning fork for absolute magnitude.

Based on magnitude-limited samples, \citet{van11} found no significant difference in luminosity distributions between unbarred and barred galaxies. It is interesting to note that for a given subclass, S(B)a through S(B)d, barred galaxies tend to be heavier in our study, as recently found by \citet{oh12} in their in-depth study on barred galaxies.  This result does not change when we use different mass estimates (e.g., NYU-VAGC \citep{bla05} and MPA-JHU catalogue\footnote{\tt{http://www.mpa-garching.mpg.de/SDSS/DR7/}}). This result may not be statistically significant as their scatters overlap based on our sample, and it calls for further investigations.

\subsection{Velocity Dispersion}

\begin{figure}
\centering
\includegraphics[width=0.5\textwidth]{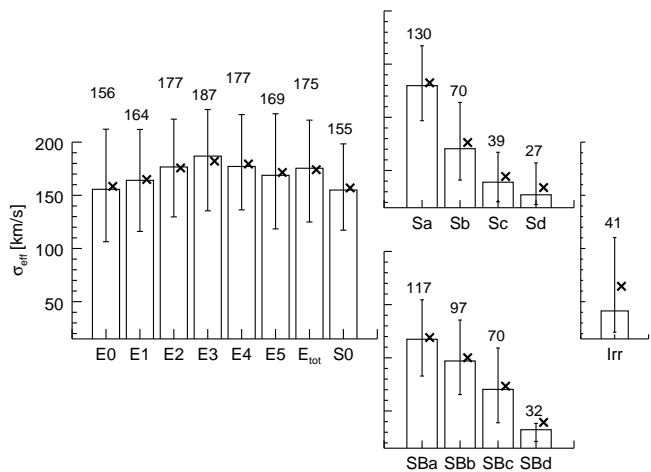}
\caption{Velocity dispersion. The format is the same as that of Fig.4.}
\label{sigma}
\end{figure}

One of the most important differences between early and late-type galaxies from textbooks is their kinematic status. Early-type galaxies are more pressure supported, whereas late-types are more rotationally supported. SDSS provides spectra and associated stellar velocity dispersions for most of its galaxies. We adopted the new measurements of velocity dispersions from the OSSY database \citep{oh11}. The OSSY team re-measured the velocity dispersions based on their improved spectral fits to the observed spectra. The difference between the SDSS pipeline values and the OSSY values was minor, but more obvious at low values (see Fig.~6 of Oh et al. 2011), which indicated that the velocity dispersions we used were significantly different from the SDSS values for spiral galaxies. We corrected the OSSY velocity dispersions further for the effective radius of each galaxy and derived $\sigma_{\rm eff}$ using the formulae from \citet{cap06} and \citet{gra05}. We adopted the values only when $10<\sigma_{\rm eff}<400$ \kms. The reason for this range is that, while velocity dispersion measurements are prone to many uncertainties, values that are smaller than this range are extremely difficult to measure realistically and values that are larger than this range are thought to be unrealistic. Roughly 10\% of S(B)c, 28\% of S(B)d and, 10\% of irregular galaxies were eliminated by this cut. Therefore, some of our final galaxy samples do not appear in this figure.

The median and mean values of sigma are shown in Fig.~\ref{sigma}. Early-type galaxies had larger velocity dispersion values ($175^{+45}_{-51}$ \kms) than late-types ($60^{+48}_{-32}$ \kms), as expected from the literature \citep{veg01, agu09}. Within early-types, our data did not show any systematic trend and the scatter was large. As has been repeatedly pointed out, sub-classification indexes (E0 through E5) may not trace much of the intrinsic properties of galaxies. Spiral galaxies do show an apparent trend, just as in the case of stellar mass in the previous section.


\section{Absorption-Line Strengths}
\label{sec:absorption}

In this section, we provide the characteristic properties of the stellar absorption-line strengths of our galaxies based on the OSSY database \citep{oh11}. The OSSY team measured the absorption-line strengths from emission-cleaned spectra with the standard line strength definition. Before measuring the strengths, the SDSS spectra were set to the rest-frame and degraded to the Lick/IDS system resolution. Also, they corrected the stellar kinematic broadening by taking into consideration the optimal combination of the stellar templates and the emission subtracted \gandalf\ fit. In addition, when the telluric lines fell into the pseudo continuum or index passbands, the line index values were replaced by the values of the stellar optimal fit in order to avoid meaningless index strengths from unacceptable fits. We used quality-assessing parameters that were smaller than 2 in the OSSY catalogue, which guaranteed the reliability of the continuum fitting process within 2$\sigma$ at all of the S/N regimes. In other words, we only used the galaxies for which the continuum fits were achieved with good confidence. By using this criterion, roughly 3--5\% of galaxies from each morphology class were eliminated, with the exception of Sd (SBd) (5--15\%) and irregulars (5--8\%). The dramatic rejection rate of Sd (SBd) and irregulars was a result of being faint and, therefore, it led to a low signal-to-noise ratio compared to other types. As a result, we presented 7,252 (97.6 \%), 7,173 (96.5 \%), and 7,204 (96.9 \%) galaxies for \Hb, Fe5270, and Mgb index strengths, respectively.

\begin{figure}
\centering
\includegraphics[width=0.5\textwidth]{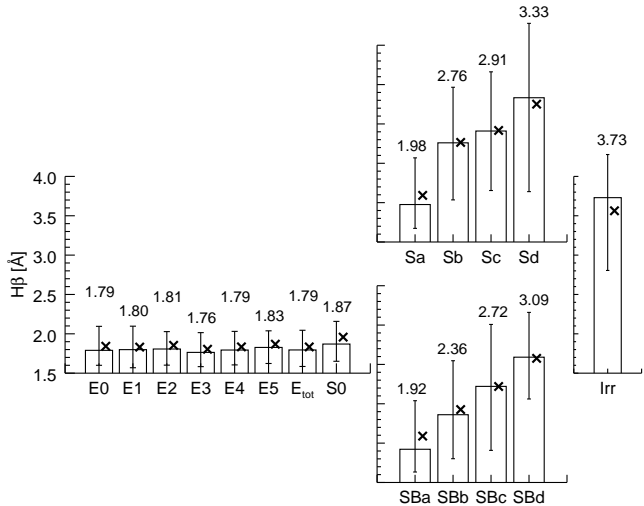}
\caption{\Hb\ absorption line strengths. The format is the same as that of Fig.4.}
\label{hbeta}
\end{figure}

\subsection{H$\beta$}
\label{sec:hbeta}

Balmer absorption-lines are widely used as a galaxy age indicator due to their high sensitivity to age, but not to metallicity \citep{wor94}. H$\beta$ is preferred in particular for its good strength and convenient wavelength. It peaks around A-type stars whose photosphere is roughly 10,000~K degrees and, therefore, it is very useful for deriving the ages of stellar populations older than 100Myr whose turn-off temperature roughly corresponds to A type. 

Fig.~\ref{hbeta} shows the equivalent width of the \Hb\ absorption-line strength. Elliptical galaxies have small \Hb\ values (median $1.79^{+0.25}_{-0.21}$ \AA), while spiral galaxies have large values ($2.77^{+0.76}_{-0.77}$ \AA). This measurement is consistent with previous studies \citep{tan98, ber06, kun06, gan07} in 1$\sigma$ range. The larger values of the late-type galaxies indicate relatively younger luminosity-weighted ages. A good trend among the late-type galaxies along the Hubble Sequence is also consistent with the general expectations. However, a more realistic interpretation requires taking the variation of metallicity into consideration as well, which will follow in Section 4.4.

\begin{figure}
\centering
\includegraphics[width=0.5\textwidth]{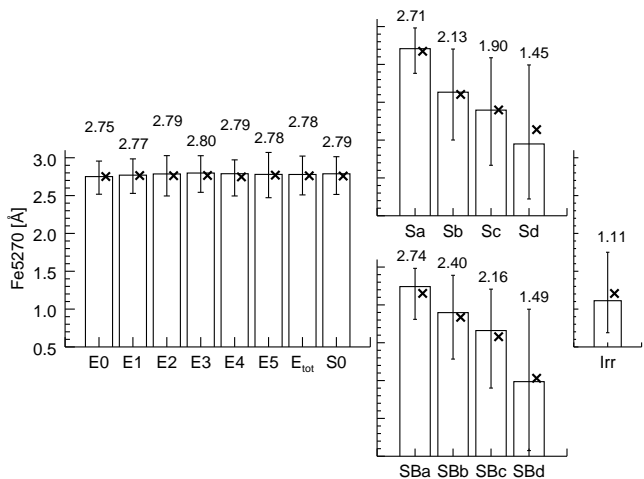}
\caption{Fe5270 absorption line strengths. The format is the same as that of Fig.4.}
\label{fe5270}
\end{figure}

\subsection{Fe5270}

The metallicity of a galaxy holds important information about its formation history. The fact that more massive galaxies are more metal rich is generally interpreted as a result of chemical recycling, which is affected by the size of the gravitational potential well of the galaxy \citep{lar74}. Elliptical galaxies are generally more massive than spirals, as discussed in Sections 3.3 and 3.4, and also believed that they have had more effective recycling of metals through shorter starbursts, and, therefore, these galaxies are expected to have a higher metal abundance.

We inspected the classical metallicity indicator, the Fe5270 index. Our cursory inspection shown on Fig.~\ref{fe5270} confirms our expectation that massive elliptical galaxies have stronger Fe5270 strengths and that the Fe5270 strength decreases among the late-types along the Hubble Sequence (from S(B)a to S(B)d). The type dependence of Fe5270 and other lines (Mgb and \Hb\ in the following sections) is partly due to the type-mass trend but still visible even for fixed mass. As mentioned in Section 4.1, however, the age and the metallicity of a galaxy both affect the Fe5270 index. Therefore, estimating the metallicity of a galaxy solely based on the Fe5270 is not feasible. We should constrain the age and metallicity parameters simultaneously instead \citep{gon93, fish96, jor99, kun00, tra00a}. We will discuss this further in Section 4.4.

\begin{figure}
\centering
\includegraphics[width=0.5\textwidth]{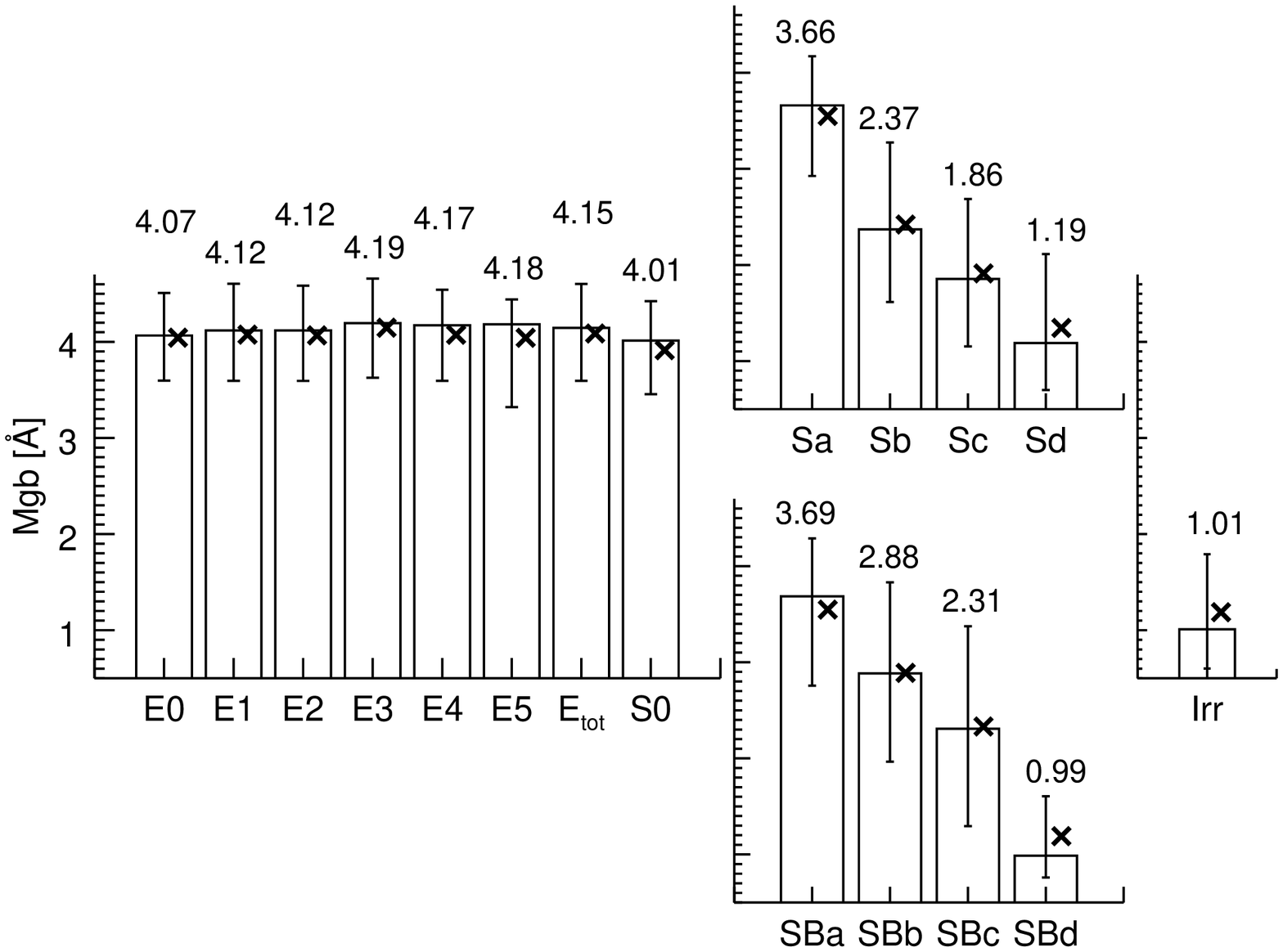}
\caption{Mgb absorption line strengths. The format is the same as that of Fig.4.}
\label{mgb}
\end{figure}

\subsection{Mgb}

The Mgb index is generally considered to be a good tracer of the $\alpha$ particle abundance. The $\alpha$-to-iron ratio, [$\alpha$/Fe], is thought to be indicative of the star formation history of the stellar population. While Fe is produced primarily by Type Ia supernova explosions, Mg, which is the most representative $\alpha$ element, is produced by both Type Ia and II supernovae \citep{tin79, nom84, woo95, thi96}. Since it takes longer for a stellar population to produce Type Ia supernovae than to produce Type II supernovae, the duration of the galactic-scale star formation history directly determines the value of [$\alpha$/Fe] (e.g., \citealt{gre83, mat86, mat87, pag95, McW97}). In this line of thought, the Hubble Sequence shown in Fig.~\ref{mgb} can be naturally explained. The Mgb index is substantially higher in early-type galaxies, which is interpreted as a result of a relatively shorter star formation timescale. Late-type galaxies on the other hand show markedly lower values of Mgb with a clear decreasing trend from the S(B)a to S(B)d along the Hubble Sequence. This result is consistent with the general understanding of a longer star formation timescale in a later Hubble type. Comparing Fig.~\ref{fe5270} with Fig.~\ref{mgb}, we can see that the Mg line strength changes more dramatically with Hubble type than Fe5270 does. As we mentioned in the previous two sections, all of the available information must be considered together in order to derive [$\alpha$/Fe] based on the Mgb measurements.

\begin{figure}
\centering
\includegraphics[width=0.5\textwidth]{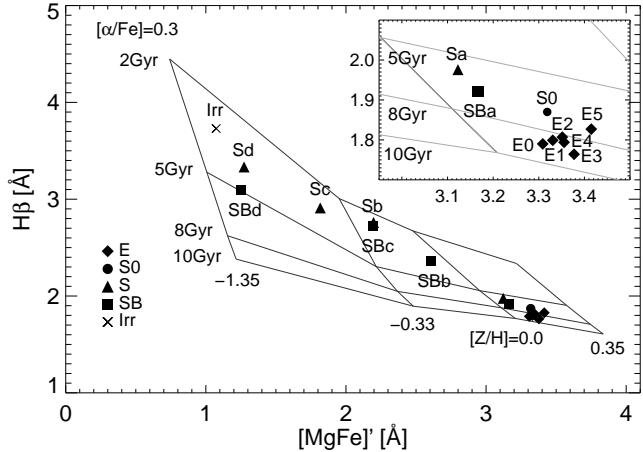}
\caption{[MgFe]' vs \Hb\ absorption indexes with model grids from \citet{tho03}. 
The median values are shown here. 
The inset zooms in on the high [MgFe]' and low \Hb\ regions. 
The standard deviation values are omitted for clarity, but can be seen in Tab.~\ref{tab:tablefour} and Tab.~\ref{tab:tablefive}.}
\label{amd}
\end{figure}

\subsection{Comparison with Stellar Population Models}

In principle, Balmer lines and metal lines are affected by both age and metallicity at some level and, therefore, the question is whether we can find a good combination of line indexes with which the degeneracy spell can be broken. It was \citet{wor94} who did this first and numerous investigations followed this approach.

Fig.~\ref{amd} shows the average (median) properties of Hubble types in comparison to the simple stellar population model grids \citep{tho03}. For our cursory inspection, we took the models of [$\alpha$/Fe] = 0.3 for simplicity. We adopted [MgFe]' as a metallicity indicator from \citet{tho03}. This index is relatively less sensitive to [$\alpha$/Fe], but still reflects the contribution of $\alpha$ elements to metallicity. It should be noted that, even in this \Ha--[MgFe]' plane which is supposed to be more successful than any other for breaking the age-metallicity degeneracy spell, the degeneracy was still visible. In other words, age and metallicity can have similar effects on the two indexes.

Roughly speaking, there was a general trend of decreasing age and metallicity from early- to late-type galaxies. Early-type galaxies were barely distinguishable in age or metallicity between subclasses and had median ages of roughly 8 Gyr and metallicities that were above Solar. However, it is important to remember that the age and metallicity estimates were the luminosity-weighted values and, therefore, ``a frosting effect'' from a small fraction (say a few percent) of young stars could easily influence them \citep{tra00a}.

\begin{figure}
\centering
\includegraphics[width=0.5\textwidth]{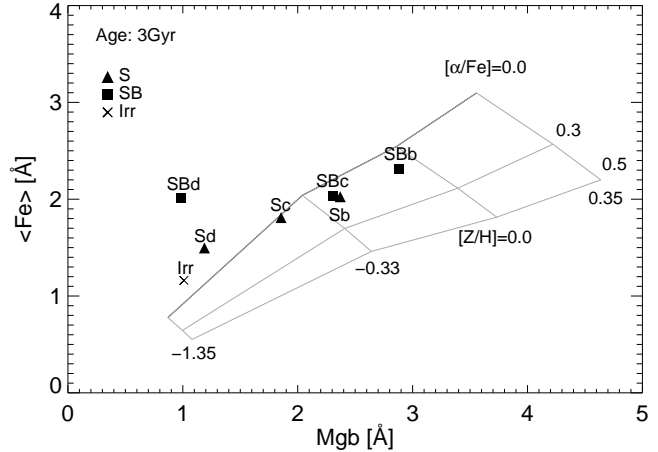}
\caption{Mgb vs $<$Fe$>$ for late-type galaxies with the 3 Gyr model grids of TMB03. 
Note that Sa and SBa galaxies are shown in Fig.13 with the early-type galaxies. See text for details.}
\label{3gyr}
\end{figure}

\begin{figure}
\centering
\includegraphics[width=0.5\textwidth]{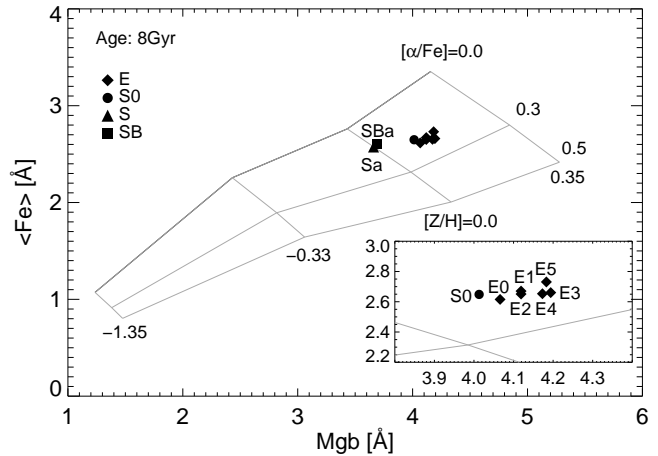}
\caption{Same as Fig.~\ref{amd} but for Sa, SBa, and early-type galaxies. The models are for the age of 8 Gyr. See text for details.}
\label{8gyr}
\end{figure}

In comparison, the luminosity-weighted ages of the spiral galaxies were substantially smaller, between 2 and 5 Gyr. It should be noted that the SDSS spectra were sampled from the SDSS fiber that covered only the central 1.5'' in radius. While the details depend on the apparent size and morphology type of the galaxy, this method basically sampled more lights from the central bulge and, therefore,  may not represent the entire galaxy. This fact should be kept in mind when it is noted that the spiral bulges from this exercise appeared to be sub-solar in metallicity. However, this result was also subject to the ``frosting effect''.

\begin{figure*}
\centering
\includegraphics[width=1\textwidth]{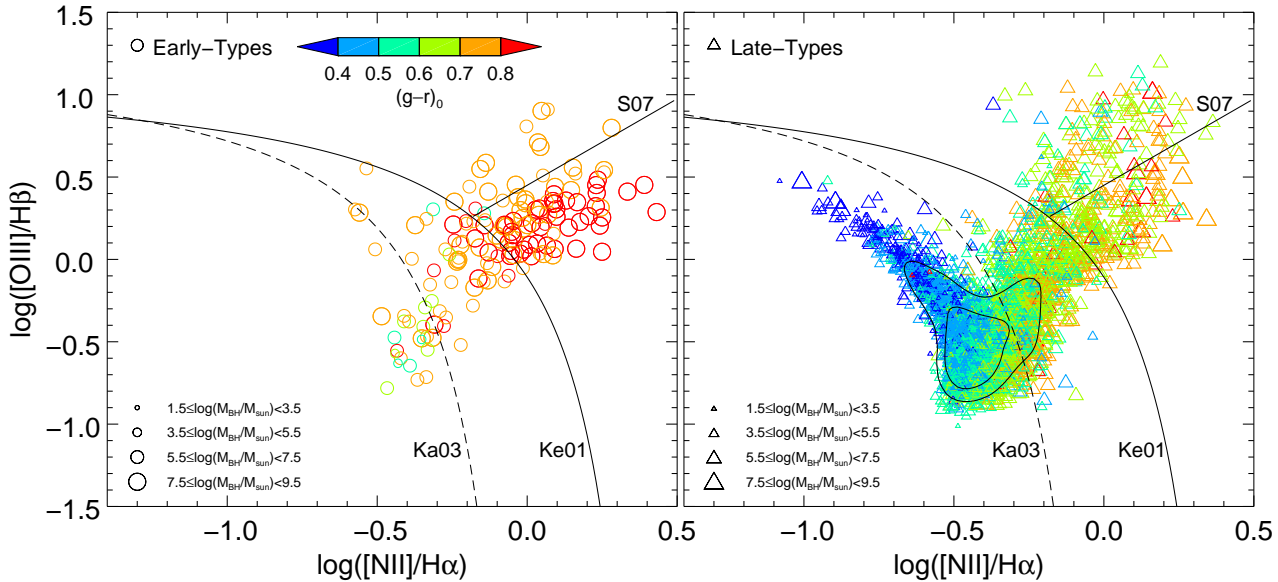}
\caption{The BPT diagnostic diagram. 
The 3,543 galaxies having $A/N \geq 3$ for $all$ four emission-lines are plotted in this figure. 
The left and right panels show the early and late-type galaxies, respectively. 
The color of the symbol indicates the k-corrected (g-r) color of the galaxy 
and the size of the symbol indicates the black hole mass estimated from the velocity dispersion. 
The dashed line is the empirical star-formation curve obtained from \citet{kau03} 
and the solid line is the theoretical maximum starburst model of \citet{kew01}. 
The solid-straight line is the empirical demarcation of \citet{sch07} distinguishing the Seyfert AGN from the LINERs. 
The inner and outer contours in the right panel indicate 0.5$\sigma$ and 1.0$\sigma$, respectively.}
\label{bpt}
\end{figure*}

As mentioned in Section 4.3, [$\alpha$/Fe] can be constrained by comparing observational data with simple stellar population models. Figs.~\ref{3gyr} \& \ref{8gyr} show the 3 Gyr and 8 Gyr model grids for the late-type and early-type galaxies, respectively. The choice of the typical age for late- (3 Gyr) and early- (8 Gyr) type galaxies was roughly based on Fig.~\ref{amd}. The Sa and SBa galaxies are shown in Fig.~\ref{8gyr} with the early-type galaxies, because their characteristics in age and metallicity (within the SDSS fiber) were more comparable to those of early-type galaxies than to those of late-type galaxies.

The $\alpha$-to-iron index, [$\alpha$/Fe], was derived from the Mgb - $<$Fe$>$ plane, where $<$Fe$>$ was defined as (Fe5270+Fe5335)/2 by \citet{gor90}. The late-type galaxies in Fig.~\ref{3gyr} were roughly consistent with [$\alpha$/Fe] = 0.0, whereas the early-type galaxies in Fig.~\ref{8gyr} favored larger values between [$\alpha$/Fe] = 0.0 and 0.3. Our results are all consistent with those of previous studies \citep{wor92, vaz97, tra00b, tho05, schi07}. This procedure was undoubtedly based on circular logic, as we first selected rough age estimates by pre-fixing the value of [$\alpha$/Fe] (0.3) in Fig.~\ref{amd}. In principle, we could overcome this circular argument by iterating this process on each Hubble type (instead of using a uniform value of [$\alpha$/Fe] for all of the types). However, it would not be very meaningful to derive properties beyond what was already done in this study, because our aim was not to derive accurate values of ages and metallicities of galaxies. Instead, our aim was to demonstrate how to use our database in order to derive such properties in more detailed application studies in the future. In addition, such an iteration would require multiple steps of interpolations with model grids, which would generate numerical errors that would not be much smaller than the uncertainty we have in our current simple analysis.

Extreme late-type galaxies (Sd, SBd) and irregular galaxies extend outside the grids in Fig.~\ref{3gyr}. This result may indicate [$\alpha$/Fe] $<$ 0.0, but it may also be due to the small number statistics.


\section{Emission-Line Statistics}
\label{sec:emission}

\begin{deluxetable*}{lccccccccccccc}
\tabletypesize{\scriptsize}
\tablecaption{Emission-line classification results}
\tablewidth{0pt}
\tablehead{
\colhead{\multirow{2}{*}{Classification}} &
\colhead{\multirow{2}{*}{N}} & 
\colhead{\%(emission)\tablenotemark{a}} & 
\colhead{\multirow{2}{*}{$E_{tot}$}}	&
\colhead{\multirow{2}{*}{S0}}	&
\colhead{\multirow{2}{*}{Sa}}	&
\colhead{\multirow{2}{*}{Sb}}	&
\colhead{\multirow{2}{*}{Sc}}	&
\colhead{\multirow{2}{*}{Sd}}	&
\colhead{\multirow{2}{*}{SBa}}	&
\colhead{\multirow{2}{*}{SBb}}	&
\colhead{\multirow{2}{*}{SBc}}	&
\colhead{\multirow{2}{*}{SBd}}	&
\colhead{\multirow{2}{*}{Irr}}	\\
\colhead{}	&
\colhead{}	&
\colhead{\%(total)\tablenotemark{b}} &
\colhead{}	&
\colhead{}	&
\colhead{}	&
\colhead{}	&
\colhead{}	&
\colhead{}	&
\colhead{}	&
\colhead{}	&
\colhead{}	&
\colhead{}	&
\colhead{}}
\startdata
\multirow{2}{*}{Emission-line galaxies\tablenotemark{c}}	&	\multirow{2}{*}{3,543}	&	100		&	3.33		&	0.93		&	2.96		&	38.87	&	30.68	&	2.09		&	1.72		&	11.01	&	5.39		&	0.40		&	2.62 \\
							&						&	(48)		&	(1.59)	&	(0.44)	&	(1.41)	&	(18.54)	&	(14.63)	&	(1.00)	&	(0.82)	&	(5.25)	&	(2.57)	&	(0.19)	&	(1.25) \\
\multirow{2}{*}{Star-forming}		&	\multirow{2}{*}{2,345}	&	66		&	0.51		&	0.14		&	0.62		&	25.23	&	26.50	&	2.09		&	0.56		&	4.15		&	3.56		&	0.40		&	2.43 \\
							&						&	(32)		&	(0.24)	&	(0.07)	&	(0.30)	&	(12.03)	&	(12.64)	&	(1.00)	&	(0.27)	&	(1.98)	&	(1.70)	&	(0.19)	&	(1.16) \\
\multirow{2}{*}{Transition region}		&	\multirow{2}{*}{775}		&	22		&	0.87		&	0.28		&	1.30		&	9.68		&	3.25		&	0.00		&	0.51		&	4.40		&	1.38		&	0.00		&	0.20 \\
							&						&	(10)		&	(0.42)	&	(0.13)	&	(0.62)	&	(4.62)	&	(1.55)	&	(0.00)	&	(0.24)	&	(2.10)	&	(0.66)	&	(0.00)	&	(0.09) \\
\multirow{2}{*}{Seyfert}			&	\multirow{2}{*}{170}		&	5		&	0.25		&	0.14		&	0.28		&	2.03		&	0.45		&	0.00		&	0.14		&	1.24		&	0.25		&	0.00		&	0.00 \\
							&						&	(2)		&	(0.12)	&	(0.07)	&	(0.13)	&	(0.97)	&	(0.22)	&	(0.00)	&	(0.07)	&	(0.59)	&	(0.12)	&	(0.00)	&	(0.00) \\
\multirow{2}{*}{LINER}			&	\multirow{2}{*}{253}		&	7		&	1.69		&	0.37		&	0.76		&	1.92		&	0.48		&	0.00		&	0.51		&	1.21		&	0.20		&	0.00		&	0.00 \\
							&						&	(3)		&	(0.81)	&	(0.17)	&	(0.36)	&	(0.92)	&	(0.23)	&	(0.00)	&	(0.24)	&	(0.58)	&	(0.09)	&	(0.00)	&	(0.00) \\

Unclear\tablenotemark{d}		&	3447	&	(46)\tablenotemark{b}		&	(12.06)	&	(3.22)	&	(3.49)	&	(10.23)	&	(10.30)	&	(0.63)	&	(1.59)	&	(3.30)	&	(1.28)	&	(0.05)	&	(0.26) \\	

Weak-emission\tablenotemark{e}	&	439		&	(6)\tablenotemark{b}			&	(4.64)	&	(0.78)	&	(0.19)	&	(0.09)	&	(0.00)	&	(0.00)	&	(0.15)	&	(0.05)	&	(0.00)	&	(0.00)	&	(0.00) \\
\enddata

\label{tab:tableone}
\tablecomments{The upper and lower rows for each morphology class indicate the percent of the ``emission-line'' galaxies and total sample, respectively.}
\tablenotetext{a}{Per cent of ``emission-line'' galaxies (N=3,543).}
\tablenotetext{b}{Per cent of the total sample (N=7,429).}
\tablenotetext{c}{$A/N \geq 3$ for \NII, \Ha, \OIII\ and \Hb\ emission lines}
\tablenotetext{d}{$1 < A/N < 3$}
\tablenotetext{e}{$A/N \leq 1$}
\end{deluxetable*}

Nebular emission lines reveal the physical state of ionized gas and, therefore, are useful for studying nuclear activities around a central supermassive black hole and star formation. It is not trivial to distinguish one effect from the other, because both star forming regions and active galactic nuclei (AGN) can excite Balmer and forbidden lines. Baldwin, Phillips \&\ Terlevich (1981, hereafter referred to as BPT) introduced several combinations of emission line ratios that can be used to do just this (see also, \citealt{kew01, kau03}). In comparison to star forming regions, AGN are assumed to produce photons with higher energy and, therefore, are more effective at producing extended partially ionized regions and inducing collisional excitations. These results lead to higher ratios of the collisionally-excited forbidden lines over the photoionization-induced Balmer emission lines. The idea has been confirmed further by \citet{vei87} and now even finer classifications of the AGN, such as Seyferts and low ionization nuclear emission line regions (LINERs), are often attempted \citep{hec80, kew06}.

\begin{figure}
\centering
\includegraphics[width=0.5\textwidth]{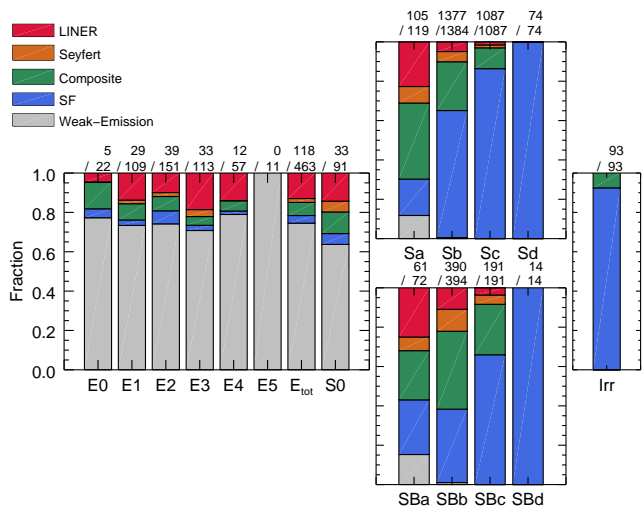}
\caption{Sub-classification in terms of emission line statistics. This figure shows the fractions of weak-emission, star-forming, composite, Seyfert AGN, and LINER host galaxies in each classification. There are two numbers above each bar. The first number indicates the number of ``emission-line'' galaxies that show all four emission lines with $A/N \geq 3$ confidence, and the second number indicates the number of ``emission-line'' galaxies plus ``weak-emission'' galaxies that have all four lines with $A/N \leq 1$. Unclear galaxies with $1 < A/N < 3$ do not appear in this plot.}
\label{pasfagn}
\end{figure}

In Fig.~\ref{bpt}, we present a BPT diagnostic diagram of our galaxies with emission lines. We selected the emission line galaxies using the cut $A/N \geq 3$, where $A$ is the best fitting amplitude of the emission line being considered and $N$ is the level of noise in the residuals of the \gandalf\ fit (see \citealt{oh11} for details). This cut selected galaxies with statistically significant emission lines on {\em all four lines} we considered in this diagnosis. Approximately 48\% of our samples exhibited all four emission lines above this cut: 9\% of the early-type galaxies and 59\% of the late-type galaxies. However, it should be noted that this does not mean that 52\% of our samples were non-emission galaxies. Many of the galaxies satisfied our A/N criterion only on two or three lines instead of all four.

In Fig.~\ref{bpt}, the size of each symbol indicates the black hole mass, while its color indicates the $(g-r)_0$ color of the galaxy. The black hole mass was derived based simply on velocity dispersion following \citet{gul09}. The emission-line classification results are summarized in Tab.~\ref{tab:tableone}.

As expected, the late-type galaxies (right panel) were optically blue and most of them showed emission lines due to star formation. Their black hole masses were generally smaller than those of the early-type galaxies. Most early-type emission-line galaxies (left panel) were classified as LINERs. Among AGNs, LINERs in particular have a heavier black hole than Seyferts do. A similar trend of increasing black hole mass from left to right is visible in the early-type galaxies as well. The results of this study were consistent with the earlier findings of \citet{sch07}. The sub-classifications are summarized in Fig.~\ref{pasfagn}. The galaxies with low or no emission lines ($A/N \leq 1$) were categorized as ``weak-emission galaxies'' and those with $1 < A/N < 3$ were classified as ``unclear'' and, therefore, they were excluded from this figure. For example, the numbers above the bar for $E_{\rm tot}$ indicate that there were 463 elliptical galaxies altogether that were either classified as ``emission-line'' or ``weak-emission''. Of the 463 elliptical galaxies, 118 were ``emission-line'' galaxies. Since there were 1,359 elliptical galaxies altogether in Fig.~\ref{sample}, then 896 (from $1359-463$) of the elliptical galaxies were classified as ``unclear''. Considering that a bulk of the galaxies were unclassifiable in terms of central activities, our statistics (fractions) should be used with caution. 

Earlier type galaxies consistently showed a higher AGN fraction. In addition, there was a clear tendency for an increasing fraction of star-forming galaxies within late-type galaxies along the Hubble Sequence. The extreme late-type galaxies, Sd, SBd, and Irr, were primarily star forming, as expected.

\begin{figure}
\centering
\includegraphics[width=0.5\textwidth]{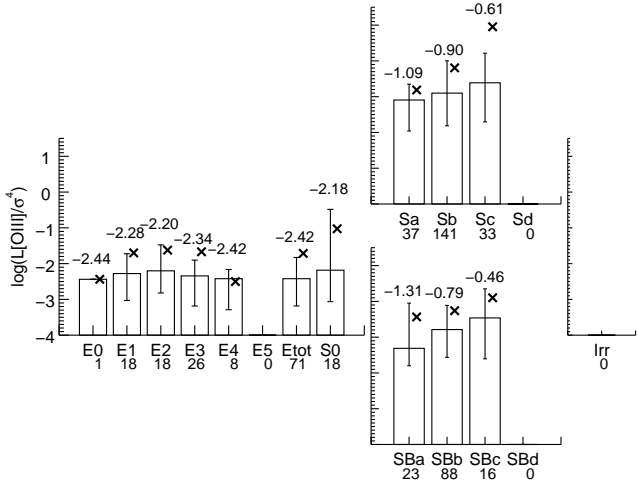}
\caption{[OIII]/$\sigma^4$ as a proxy to the black hole mass accretion rate for the AGN host galaxies. 
The number below each classification label indicates the number of galaxies in each bin.}
\label{oiii}
\end{figure}

\subsection{Accretion Rate}

The forbidden emission line from doubly ionized oxygen, [OIII], is often used to estimate the black hole gas accretion rate \citep{kau03}. The ionization near AGN is caused by the central black hole and, therefore, a strong [OIII] emission indicates a high accretion rate by the black hole. In order to eliminate the effect of the black hole mass influencing the emission-line strengths, we divided the line strength from the OSSY catalogue by $\sigma^4$ \citep{hec04, bes05, kew06}, assuming that the [OIII] luminosity scales with the AGN bolometric luminosity and $\sigma^4$ traces the mass of the black hole \citep{fer00, geb00}. Fig.~\ref{oiii} shows the characteristic values for Hubble classes. Elliptical galaxies have a larger black hole than late-type galaxies and, therefore, they have a stronger [OIII] emission line. However, when normalized by the proxy to the black hole mass, the elliptical galaxies do in fact show a lower value of accretion rate. This result is consistent with earlier works by \citet{hec04} and \citet{wu04}. This result is perhaps due to the fact that elliptical galaxies have a less copious supply of gas for accretion.

An important caveat of using the SDSS data is that SDSS uses a 3" diameter fixed fiber. \citet{sar10} noted that such a wide fiber will allow much of the diffuse stellar light to contaminate the lights from the black hole gas accretion. As a result, many of the LINER candidates based on the SDSS spectra may not be real AGN.

\begin{figure}
\centering
\includegraphics[width=0.5\textwidth]{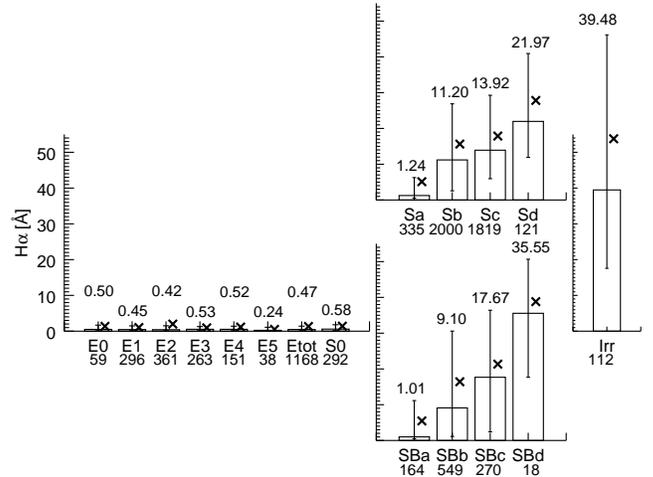}
\caption{The \Ha\ equivalent width of star-forming galaxies. 
The number below each classification label indicates the number of galaxies in each bin.}
\label{halpha}
\end{figure}

\subsection{$H\alpha$ Emission Line}

Hydrogen emission lines have been widely used as an indicator of star formation \citep{coh76, ken83, rom90, gav91, ken92, ryd94, gal95, ken98, bri04, sal07}. Young, hot stars produce HII regions first and then the ionized hydrogens undergo the recombination process, thereby resulting in the emitting of line fluxes. We show  in Fig.~\ref{halpha} the $H\alpha$ emission line strength in equivalent width from the OSSY catalogue, along the Hubble Sequence. We show only the non-AGN host galaxy candidates based on the BPT diagnostics. The emission line strength normalized by the stellar mass (because it was given in equivalent width) was extremely low in early-type galaxies. There was a clear trend of increasing emission line strength along the Hubble Sequence among late-type galaxies.


\begin{figure}
\centering
\includegraphics[width=0.5\textwidth]{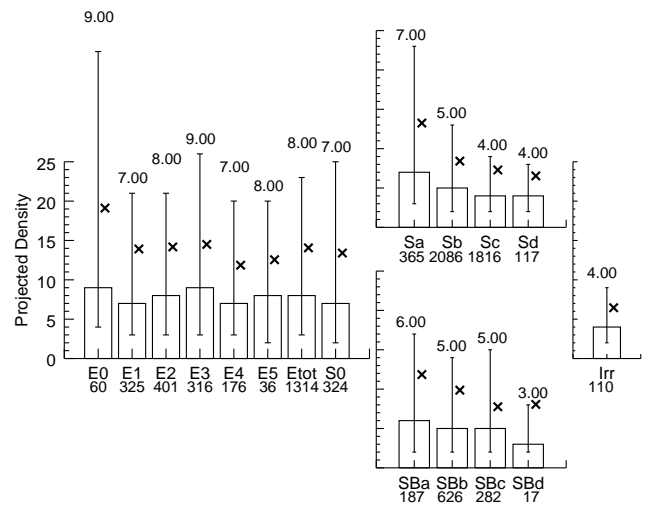}
\caption{The number of galaxies inside a search cylinder. This tuning fork diagram shows how many galaxies with the same luminosity cut ($M_{\rm r} \leq -18.65$) were found to in our sample redshift range and within the 25 arcsec projected radius. The median values are shown above each bar. Note that the numbers of the galaxies shown below the bars are somewhat different from those in Fig. 2, because the 185 galaxies close to the SDSS field boundary have been removed for consistent number counting (see text).}
\label{numden}
\end{figure}

\section{Density}
\label{sec:density}

In this section, we discuss whether or not there is any trend or peculiarity in the Hubble Sequence in terms of the local environment. By simply counting the galaxies that were within 25 arcsec (roughly 1~ Mpc  at the distance of our sample galaxies) from our target galaxy, we measured its projected density parameter. Both the target and neighboring galaxies were required to satisfy the same luminosity cut, $M_{\rm r} \leq -18.65$. We chose this cylindrical approach over elaborate sphere or ellipsoid approaches, because it is difficult to consistently correct the peculiar motion effect along the line of sight through all of the density ranges, no matter which shape is chosen for the search volume. If our target galaxies were all inside clusters, then an ellipsoidal search would have been more effective \citep{yoo08}. However, our samples were heavily mixed in terms of the environment. We tried giving weights based on distance \citep{yoo08}, but it only caused a minor difference in our results. Therefore, we decided to simplify our density measurement so that its meaning (as a density measure) was clearer.

A density measurement would be inaccurate for the galaxies near the SDSS survey boundary, because their search cylinder cannot be complete. Therefore, we removed 185 (2.5\%) such galaxies from our study samples. It should be noted that we only counted the galaxies with spectroscopic information and, therefore, our density parameter was subject to the spectroscopic completeness of the SDSS survey \citep{yoo08}. It should also be noted that we were counting all the galaxies above the brightness cut in the redshift range, whether they were morphologically classified or not, as we did not make any distinction in the morphology when counting the galaxies.

Fig.~\ref{numden} shows our results for density. The median of the density parameters was 5 for the entire sample. This result indicates that there were typically only 5 SDSS spectroscopic galaxies with $M_{\rm r} \leq -18.65$ inside a search cylinder. Quite a few galaxies (823, 11\%) had a density parameter greater than 20 and 42 of them even reached beyond 100. Visual inspection of their SDSS images indicated that they were likely associated with galaxy clusters. We did indeed identify the following 15 Abell clusters overlapping with our sample: Abell 2063, 1314, 1228, 1142, 2147, 2151, 2152, 1139, 2593, 2506, 1257, 2052, 2107, 119, and 957. The early-type galaxies showed a markedly higher value than the late-type galaxies. In particular, the median and mean values for the giant elliptical galaxies with $M_{\rm r} \leq -22$ were 11 and 17.28, respectively, suggesting heavy clustering. The result from this simplistic test is qualitatively consistent with the well-known morphology-density relation of \citet{dre80}. The trend in density along the Hubble Sequence, if real, would be sensible based on the hierarchical merger paradigm \citep{spr01} and the local density effect on morphology (e.g., \citealt{got03, tan04, van08, kaw08}).

\section{Summary}
\label{sec:summary}

We presented the statistical average properties of galaxies along the Hubble Sequence based on a volume-limited sample of 7,429 morphologically classified galaxies from the SDSS DR7 database. The galaxies were drawn from the narrow redshift range of $0.033 < z < 0.044$ so that the evolution effect could be ignored. Our samples reached down to the brightness of $M_{\rm r} = -18.65$ and, therefore, covered much fainter galaxies than many of the previous studies. The novelty of our study is that we adopted the recent spectroscopic database from the OSSY catalogue \citep{oh11}. We summarized some of the notable features of our results below. First, it should be noted that the properties based on spectroscopic data were for the central region of the galaxy covered by the 1.5 arcsec radius fiber of SDSS.

\begin{itemize}
  \item Our sample was severely biased against apparently small and low $IsoB_r$/$IsoA_r$ galaxies due to the difficulty in morphology classification and, therefore, does not properly represent the E5, E6, E7, S0, Sd, SBd, and Irr galaxies.
  \item The distribution (in percent) was roughly ellipticals : lenticulars : unbarred spirals: barred spirals: irregulars = 18.3: 4.4: 60.5: 15.3: 1.5. This result is in good agreement with the results of recent studies \citep{nai10, de11}. The dominance of late-type galaxies was probably caused by the faint magnitude limit of our sample.
   \item In the color-magnitude diagram, galaxies with different morphology (elliptical, unbarred spiral, and barred spiral)
seemed to have a preferred region. In particular, the barred spiral galaxies were, on average, much brighter and redder than unbarred spiral galaxies. This result is consistent with the recent findings of \citet{oh12}.
   \item The stellar mass in spiral galaxies decreased from S(B)a to S(B)d regardless of the bar features.
   \item The elliptical galaxies had an average velocity dispersion of 175 \kms, which is 3 times larger than that of spiral galaxies (60 \kms).
   \item The age indicator, \Hb\ absorption-line strength, increased from S(B)a (2.0\AA) to S(B)d (3.2\AA) in spiral galaxies, while it was much lower (1.8\AA) in early-type galaxies.
   \item The metallicity indicator, Fe5270 absorption line strength, decreased from S(B)a (2.7\AA) to S(B)d (1.5\AA) among spiral galaxies, while it was higher (2.8\AA) in early-type galaxies.
  \item The Mgb absorption line strengths suggested that only early-type galaxies  were enhanced in terms of $\alpha$ element abundance over iron group abundance. This result is consistent with the results of earlier works \citep{tho10}.
  \item The use of multiple indexes compared with population models suggested that the luminosity-weighted mean ages were approximately 8 and 3Gyr for early- and late-type galaxies, respectively.
  \item Roughly half (48\%) of the entire samples showed statistically significant emission-lines for all four lines in the AGN diagnostic method we adopted. Nine percent of the early-type galaxies and 59\% of the late-type galaxies were classified as ``emission-line'' galaxies.
  \item The dominant process for the optical emission lines appeared to be AGN/LINER activity in early-type galaxies and star formation in late-type galaxies.
  \item The late-type galaxies had higher accretion rates for gas by black holes than the early-types did. This result was probably due to the fact that gas is more abundant in late-type galaxies.
  \item The \Ha\ emission line strength, a proxy to the star formation rate, was much lower in early-type galaxies. This result is consistent with previous findings \citep{ken83}.
  \item Early-type galaxies tended to reside in denser regions.
\end{itemize}
  
All of the indexes and parameters with regards to the galaxy morphology that are presented in this paper are summarized in Tab.~\ref{tab:tabletwo}, Tab.~\ref{tab:tablethree}, Tab.~\ref{tab:tablefour} and Tab.~\ref{tab:tablefive}. We are pleased to release our morphology catalogue including various spectroscopic and photometric properties (Tab.~\ref{tab:tablesix}). It is our wish that this database will be useful as a reference.

\begin{deluxetable*}{lcccccccc}
\tabletypesize{\scriptsize}
\tablecaption{Basic properties of the Early-type and Lenticular galaxies}
\tablewidth{0pt}
\tablehead{
\colhead{} &
\colhead{E0} & 
\colhead{E1} & 
\colhead{E2}	&
\colhead{E3}	&
\colhead{E4}	&
\colhead{E5}	&
\colhead{$E_{tot}$}	&
\colhead{S0}}

\startdata

Number\tablenotemark{a} & 63    & 342  & 412 & 322 & 181 & 39 & 1359 & 330  \\

&&&&&&&&\\
Cr\tablenotemark{b} & 3.20 & 3.17 & 3.22 & 3.23 & 3.20 & 3.18 & 3.21 & 3.10 \\
	
& $^{+0.19}_{-0.22}$ & $^{+0.19}_{-0.20}$ & $^{+0.15}_{-0.19}$ & $^{+0.15}_{-0.20}$ & $^{+0.19}_{-0.18}$ & $^{+0.16}_{-0.07}$ & $^{+0.17}_{-0.20}$ & $^{+0.18}_{-0.21}$ \\
	
& (3.18) & (3.16) & (3.19) & (3.20) & (3.20) & (3.22) & (3.19) & (3.08)  \\

&&&&&&&& \\	 
\fracDeV\tablenotemark{c} & 1.00 & 1.00 & 1.00 & 1.00 & 1.00 & 1.00 & 1.00 & 1.00  \\
	
& $^{+0.00}_{-0.06}$ & $^{+0.00}_{-0.06}$ & $^{+0.00}_{-0.04}$ & $^{+0.00}_{-0.05}$ & $^{+0.00}_{-0.06}$ & $^{+0.00}_{-0.06}$ & $^{+0.00}_{-0.05}$ & $^{+0.00}_{-0.16}$ \\
	
& (0.98) & (0.97) & (0.97) & (0.97) & (0.98) & (0.97) & (0.97) & (0.92)  \\

&&&&&&&& \\	
$(g-r)_0$\tablenotemark{d} & 0.80 & 0.79 & 0.79 & 0.79 & 0.79 & 0.77 & 0.79 & 0.79  \\
	
& $^{+0.02}_{-0.05}$ & $^{+0.04}_{-0.04}$ & $^{+0.04}_{-0.04}$ & $^{+0.04}_{-0.04}$ & $^{+0.04}_{-0.04}$ & $^{+0.05}_{-0.04}$ & $^{+0.04}_{-0.04}$ & $^{+0.04}_{-0.03}$ \\
	
& (0.80) & (0.79) & (0.79) & (0.79) & (0.79) & (0.79) & (0.79) & (0.79)  \\

&&&&&&&& \\	
$M_r$\tablenotemark{e} & -21.37 & -21.35 & -21.45 & -21.47 & -21.48 & -21.47 & -21.44 & -20.97  \\
	
& $^{+0.69}_{-0.56}$ & $^{+0.70}_{-0.63}$ & $^{+0.70}_{-0.58}$ & $^{+0.70}_{-0.62}$ & $^{+0.73}_{-0.52}$ & $^{+0.85}_{-0.53}$ & $^{+0.73}_{-0.59}$ & $^{+0.48}_{-0.48}$\\
	
& (-21.27) & (-21.32) & (-21.40) & (-21.44) & (-21.42) & (-21.38) & (-21.39) & (-20.97) \\

&&&&&&&& \\	
log($M_{\ast}$/$M_{\odot}$)\tablenotemark{f} & 10.99 & 10.97 & 11.02 & 11.04 & 11.01 & 10.98 & 11.00 & 10.81 \\
	
& $^{+0.26}_{-0.31}$ & $^{+0.26}_{-0.29}$ & $^{+0.23}_{-0.30}$ & $^{+0.26}_{-0.33}$ & $^{+0.25}_{-0.26}$ & $^{+0.26}_{-0.26}$ & $^{+0.27}_{-0.30}$ & $^{+0.21}_{-0.19}$\\
	
& (11.01) & (11.04) & (11.07) & (11.10) & (11.09) & (11.07) & (11.07) & (10.88)  \\

&&&&&&&& \\	
Density\tablenotemark{g} & 9.00 & 7.00 & 8.00 & 9.00 & 7.00 & 8.00 & 8.00 & 7.00\\

& $^{+30.01}_{-5.00}$ & $^{+14.00}_{-4.00}$ & $^{+13.00}_{-5.00}$ & $^{+17.00}_{-6.00}$ & $^{+13.00}_{-4.00}$ & $^{+12.00}_{-6.00}$ & $^{+15.00}_{-5.00}$ & $^{+18.00}_{-5.00}$  \\

& (19.12) & (13.92) & (14.18) & (14.50) & (11.86) & (12.56) & (14.06) & (13.40) \\

\enddata

\tablenotetext{a}{Number of galaxies for each morphology.}
\tablenotetext{b}{Concentration index, C$_r$=PetroR90/PetroR50. The rows from the top to bottom of each column represent the median, 1 $\sigma$ distribution and mean values in parenthesis.}
\tablenotetext{c}{\fracDeV, de Vaucouleurs fraction of the SDSS-r band.}
\tablenotetext{d}{k-corrected (g-r) color using the petrosian magnitude.}
\tablenotetext{e}{r-band absolute magnitude.}
\tablenotetext{f}{Logarithmic scale of the stellar mass derived from the method developed by \citet{bel01}.}
\tablenotetext{g}{Number density. Number of galaxies within a cylinder with 1 Mpc radius and $0.033 < z < 0.044$ height. 
Note that the faint galaxies were already eliminated by the volume-limitation process. The galaxies with no spectroscopic data were also not counted.}
\label{tab:tabletwo}
\end{deluxetable*}

\begin{deluxetable*}{lccccccccc}
\tabletypesize{\scriptsize}
\tablecaption{Basic properties of the Late-type and Irregular galaxies}
\tablewidth{0pt}
\tablehead{
\colhead{} &
\colhead{Sa}	&
\colhead{Sb}	&
\colhead{Sc}	&
\colhead{Sd}	&
\colhead{SBa}	&
\colhead{SBb}	&
\colhead{SBc}	&
\colhead{SBd} &
\colhead{Irr}}

\startdata

Number & 378 & 2144 & 1852 & 121 & 190 & 639 & 286 & 18 & 112 \\

&&&&&&&&& \\
Cr &  2.97 & 2.33 & 2.09 & 2.05 & 2.73 & 2.41 & 2.21 & 2.08 & 2.25 \\
	
& $^{+0.27}_{-0.37}$ & $^{+0.42}_{-0.29}$ & $^{+0.27}_{-0.16}$ & $^{+0.16}_{-0.14}$ & $^{+0.41}_{-0.49}$ & $^{+0.44}_{-0.36}$ & $^{+0.33}_{-0.25}$ & $^{+0.26}_{-0.10}$ & $^{+0.46}_{-0.23}$ \\
	
& (2.93) & (2.38) & (2.14) & (2.08) & (2.71) & (2.45) & (2.25) & (2.12) & (2.33) \\

&&&&&&&&& \\	 
\fracDeV & 0.95 & 0.24 & 0.14 & 0.14 & 0.92 & 0.73 & 0.41 & 0.44 & 0.20 \\	

& $^{+0.05}_{-0.31}$ & $^{+0.54}_{-0.23}$ & $^{+0.25}_{-0.14}$ & $^{+0.20}_{-0.14}$ & $^{+0.08}_{-0.25}$ & $^{+0.27}_{-0.46}$ & $^{+0.46}_{-0.29}$ & $^{+0.37}_{-0.33}$ & $^{+0.32}_{-0.17}$ \\
	
& (0.84) & (0.36) & (0.21) & (0.18) & (0.85) & (0.66) & (0.46) & (0.43) & (0.28) \\

&&&&&&&&&\\	
$(g-r)_0$ & 0.77 & 0.62 & 0.51 & 0.41 & 0.77 & 0.68 & 0.59 & 0.46 & 0.42 \\
	
& $^{+0.05}_{-0.04}$ & $^{+0.10}_{-0.11}$ & $^{+0.09}_{-0.08}$ & $^{+0.09}_{-0.06}$ & $^{+0.05}_{-0.07}$ & $^{+0.06}_{-0.08}$ & $^{+0.08}_{-0.13}$ & $^{+0.11}_{-0.06}$ & $^{+0.11}_{-0.09}$ \\
	
& (0.77) & (0.62) & (0.51) & (0.43) & (0.76) & (0.68) & (0.57) & (0.47) & (0.43) \\

&&&&&&&&& \\	
$M_r$ & -21.19 & -20.83 & -20.39 & -19.46 & -21.11 & -21.25 & -21.07 & -19.48 & -20.00 \\	

& $^{+0.48}_{-0.53}$ & $^{+0.58}_{-0.58}$ & $^{+0.67}_{-0.77}$ & $^{+0.39}_{-0.50}$ & $^{+0.49}_{-0.54}$ & $^{+0.60}_{-0.57}$ & $^{+1.00}_{-0.63}$ & $^{+0.45}_{-0.61}$ & $^{+0.69}_{-0.69}$ \\
	
& (-21.21) & (-20.84) & (-20.44) & (-19.53) & (-21.14) & (-21.23) & (-20.97) & (-19.63) & (-20.05) \\

&&&&&&&&& \\	
log($M_{\ast}$/$M_{\odot}$) & 10.89 & 10.57 & 10.28 & 9.83 & 10.86 & 10.81 & 10.66 & 9.96 & 9.98 \\
	
& $^{+0.22}_{-0.21}$ & $^{+0.30}_{-0.31}$ & $^{+0.38}_{-0.33}$ & $^{+0.21}_{-0.19}$ & $^{+0.21}_{-0.22}$ & $^{+0.27}_{-0.27}$ & $^{+0.29}_{-0.56}$ & $^{+0.17}_{-0.17}$ & $^{+0.40}_{-0.23}$ \\
	
& (10.95) & (10.68) & (10.43) & ( 9.89) & (10.92) & (10.88) & (10.73) & (9.96) & (10.21) \\

&&&&&&&&& \\	
Density & 7.00 & 5.00 & 4.00 & 4.00 & 6.00 & 5.00 & 5.00 & 3.00 & 4.00 \\
& $^{+16.00}_{-4.00}$ & $^{+8.00}_{-3.00}$ & $^{+5.00}_{-2.00}$ & $^{+4.00}_{-2.00}$ & $^{+11.00}_{-4.00}$ & $^{+9.00}_{-3.00}$ & $^{+10.00}_{-3.00}$ & $^{+5.00}_{-1.00}$ & $^{+5.00}_{-2.00}$ \\

&(13.25) & (8.42) & (7.29) & (6.54) & (11.85) & (9.87) & (7.77) & (8.06) & (6.45) \\

\enddata

\label{tab:tablethree}
\end{deluxetable*}

\begin{deluxetable*}{lcccccccc}
\tabletypesize{\scriptsize}
\tablecaption{Spectroscopic properties of the Early-type and Lenticular galaxies}
\tablewidth{400pt}
\tablehead{
\colhead{} &
\colhead{E0} & 
\colhead{E1} & 
\colhead{E2}	&
\colhead{E3}	&
\colhead{E4}	&
\colhead{E5}	&
\colhead{$E_{tot}$}	&
\colhead{S0}}

\startdata

Number & 63    & 342  & 412 & 322 & 181 & 39 & 1359 & 330 \\

&&&&&&&& \\	 
$\sigma_{eff}$\tablenotemark{a} & 155.7 & 164.1 & 176.7 & 186.9 & 177.1 & 168.8 & 175.4 & 155.0 \\
	
$[$\kms$]$	& $^{+56.50}_{-49.27}$ & $^{+47.83}_{-48.06}$ & $^{+44.95}_{-46.95}$ & $^{+43.81}_{-51.42}$ & $^{+48.77}_{-40.81}$ & $^{+57.88}_{-50.35}$ & $^{+45.22}_{-50.52}$ & $^{+43.40}_{-37.75}$\\
	
& (158.4) & (164.9) & (175.8) & (182.1) & (179.4) & (171.5) & (174.1) & (157.0)  \\	

&&&&&&&& \\
\Hb\tablenotemark{b} & 1.79 & 1.80 & 1.81 & 1.76 & 1.79 & 1.83 & 1.79 & 1.87\\
	
$[$\AA$]$& $^{+0.30}_{-0.19}$ & $^{+0.30}_{-0.23}$ & $^{+0.22}_{-0.20}$ & $^{+0.25}_{-0.18}$ & $^{+0.24}_{-0.19}$ & $^{+0.21}_{-0.20}$ & $^{+0.25}_{-0.21}$ & $^{+0.29}_{-0.22}$ \\
	
& (1.84) & (1.83) & (1.85) & (1.80) & (1.83) & (1.87) & (1.83) & (1.96) \\ 

&&&&&&&& \\	
Fe5270\tablenotemark{c} & 2.75 & 2.77 & 2.79 & 2.80 & 2.79 & 2.78 & 2.78 & 2.79  \\
	
$[$\AA$]$& $^{+0.20}_{-0.23}$ & $^{+0.21}_{-0.24}$ & $^{+0.24}_{-0.29}$ & $^{+0.23}_{-0.26}$ & $^{+0.18}_{-0.29}$ & $^{+0.29}_{-0.31}$ & $^{+0.24}_{-0.27}$ & $^{+0.23}_{-0.27}$ \\
	
& (2.75) & (2.76) & (2.76) & (2.77) & (2.75) & (2.77) & (2.76) & (2.76)  \\

&&&&&&&& \\	
$<$Fe$>$\tablenotemark{d} & 2.62 & 2.67 & 2.65 & 2.66 & 2.65 & 2.73 & 2.66 & 2.65  \\
	
$[$\AA$]$& $^{+0.21}_{-0.16}$ & $^{+0.22}_{-0.24}$ & $^{+0.22}_{-0.26}$ & $^{+0.22}_{-0.25}$ & $^{+0.19}_{-0.23}$ & $^{+0.15}_{-0.38}$ & $^{+0.21}_{-0.25}$ & $^{+0.22}_{-0.24}$ \\
	
& (2.61) & (2.65) & (2.64) & (2.65) & (2.63) & (2.69) & (2.64) & (2.62) \\

&&&&&&&& \\	
$[$MgFe$]$'\tablenotemark{e} & 3.31 & 3.33 & 3.35 & 3.38 & 3.36 & 3.42 & 3.35 & 3.32 \\
	
$[$\AA$]$& $^{+0.22}_{-0.28}$ & $^{+0.28}_{-0.28}$ & $^{+0.26}_{-0.36}$ & $^{+0.27}_{-0.35}$ & $^{+0.26}_{-0.38}$ & $^{+0.17}_{-0.52}$ & $^{+0.26}_{-0.33}$ & $^{+0.22}_{-0.38}$ \\
	
& (3.28) & (3.31) & (3.30) & (3.34) & (3.30) & (3.31) & (3.31) & (3.23)\\

&&&&&&&&\\	
Mgb\tablenotemark{f} & 4.07 & 4.12 & 4.12 & 4.19 & 4.17 & 4.18 & 4.15 & 4.01 \\
	
$[$\AA$]$& $^{+0.44}_{-0.47}$ & $^{+0.48}_{-0.52}$ & $^{+0.47}_{-0.53}$ & $^{+0.46}_{-0.57}$ & $^{+0.37}_{-0.58}$ & $^{+0.26}_{-0.86}$ & $^{+0.46}_{-0.55}$ & $^{+0.41}_{-0.56}$ \\
	
& (4.04) & (4.08) & (4.07) & (4.15) & (4.07) & (4.04) & (4.09) & (3.91)\\ 

&&&&&&&& \\
log($M_{BH}$/$M_{\odot}$)\tablenotemark{g} & 8.24 & 7.87 & 7.80 & 8.11 & 8.29 & 0.00 & 7.96 & 7.95 \\

(AGN) & $^{+0.00}_{-0.00}$ & $^{+0.33}_{-0.42}$ & $^{+0.51}_{-0.16}$ & $^{+0.33}_{-0.36}$ & $^{+0.25}_{-0.51}$ & $^{+0.00}_{-0.00}$ & $^{+0.39}_{-0.32}$ & $^{+0.40}_{-0.77}$ \\

& (8.24) & (7.90) & (8.02) & (8.19) & (8.25) & (0.00) & (8.10) & (8.01) \\

&&&&&&&& \\
log(L[$\mbox{O{\sc iii}}$]/$\sigma^4$)\tablenotemark{h} & -2.44 & -2.28 & -2.20 & -2.34 & -2.42 & 0.00 & -2.42 & -2.18  \\

& $^{+0.00}_{-0.00}$ & $^{+0.55}_{-0.75}$ & $^{+0.73}_{-0.62}$ & $^{+0.44}_{-0.84}$ & $^{+0.26}_{-0.87}$ & $^{+0.00}_{-0.00}$ & $^{+0.59}_{-0.76}$ & $^{+1.90}_{-0.88}$\\

& (-2.44) & (-1.70) & (-1.62) & (-1.67) & (-2.50) & (0.00) & (-1.71) & (-1.03)\\

&&&&&&&&\\	
H$\alpha$\tablenotemark{i} & 0.50 & 0.45 & 0.42 & 0.53 & 0.52 & 0.24 & 0.47 & 0.58 \\
	
(Non-AGN)$[$\AA$]$ & $^{+1.19}_{-0.39}$ & $^{+1.04}_{-0.34}$ & $^{+1.09}_{-0.30}$ & $^{+0.75}_{-0.40}$ & $^{+0.87}_{-0.38}$ & $^{+0.86}_{-0.13}$ & $^{+0.93}_{-0.35}$ & $^{+1.22}_{-0.42}$ \\
	
& (1.37) & (1.03) & (1.98) & (0.94) & (1.14) & (0.50) & (1.32) & (1.40) \\
\enddata

\label{tab:tablefour}

\tablenotetext{a}{Effective velocity dispersion calculated from the formulae by \citet{gra05} and \citet{cap06}.}
\tablenotetext{b}{\Hb\ equivalent width provided by the OSSY catalogue with the quality-assessing parameter, N$\sigma$ $<$ 2 and EW $>$ 0. The rest of the stellar absorption-line features follow the same criteria. Here, 97.7$\%$ of the sample galaxies were used by the selection cut.}
\tablenotetext{c}{Fe5270 equivalent width. 96.8$\%$ of the sample galaxies were used.}
\tablenotetext{d}{$<$Fe$>$ equivalent width. $<$Fe$>$=(Fe5270+Fe5335)/2 \citep{gor90}. 94.9$\%$ of the sample galaxies were used.}
\tablenotetext{e}{$[$MgFe$]$' equivalent width. $[$MgFe$]$'=$\sqrt{Mgb(0.72 \times Fe5270 + 0.28 \times Fe5335)}$ \citep{tho03}. 94.4$\%$ of sample galaxies are used.}
\tablenotetext{f}{Mgb equivalent width. 97.2$\%$ of the sample galaxies were used.}
\tablenotetext{g}{Central black hole mass of the AGN derived by \citet{gul09} using velocity dispersion.}
\tablenotetext{h}{L[\mbox{O{\sc iii}}] for the AGN divided by the effective velocity dispersion to the power of four. Note that the indices for the several of the morphology classes (E0, E5, Sd, SBd and Irr) were not reliable due to the sample size. The number of galaxies used to derive the index for each classification are shown in Fig.~\ref{oiii}.}
\tablenotetext{i}{H$\alpha$ equivalent width for non-AGN galaxies. The number of galaxies for each classification are shown in Fig.~\ref{halpha}.}
\end{deluxetable*}

\begin{deluxetable*}{lccccccccc}
\tabletypesize{\scriptsize}
\tablecaption{Spectroscopic properties the of the Late-type and Irregular galaxies}
\tablewidth{400pt}
\tablehead{
\colhead{} &
\colhead{Sa}	&
\colhead{Sb}	&
\colhead{Sc}	&
\colhead{Sd}	&
\colhead{SBa}	&
\colhead{SBb}	&
\colhead{SBc}	&
\colhead{SBd} &
\colhead{Irr}}

\startdata

Number &  378 & 2144 & 1852 & 121 & 190 & 639 & 286 & 18 & 112 \\

&&&&&&&&& \\	 
$\sigma_{eff}$ & 129.8 & 70.5 & 39.0 & 27.2 & 117.4 & 97.0 & 70.3 & 32.4 & 41.3 \\
	
$[$\kms$]$	& $^{+37.50}_{-32.98}$ & $^{+43.54}_{-29.51}$ & $^{+28.12}_{-18.38}$ & $^{+29.97}_{-9.13}$ & $^{+37.07}_{-34.60}$ & $^{+38.49}_{-31.57}$ & $^{+38.92}_{-31.42}$ & $^{+ 5.89}_{-11.11}$ & $^{+68.97}_{-20.10}$ \\
	
& (132.5) & (76.3) & (44.4) & (33.9) & (119.0) & (100.1) & (73.3) & (39.3) & (64.5) \\	

&&&&&&&&& \\
\Hb & 1.98 & 2.76 & 2.91 & 3.33 & 1.92 & 2.36 & 2.72 & 3.09 & 3.73 \\
	
$[$\AA$]$& $^{+0.59}_{-0.30}$ & $^{+0.71}_{-0.72}$ & $^{+0.75}_{-0.76}$ & $^{+0.94}_{-1.19}$ & $^{+0.62}_{-0.29}$ & $^{+0.69}_{-0.56}$ & $^{+0.79}_{-0.81}$ & $^{+0.57}_{-0.53}$ & $^{+0.55}_{-0.93}$ \\
	
& (2.09) & (2.77) & (2.92) & (3.25) & (2.09) & (2.43) & (2.72) & (3.08) & (3.56) \\ 

&&&&&&&&&\\	
Fe5270 & 2.71 & 2.13 & 1.90 & 1.45 & 2.74 & 2.40 & 2.16 & 1.49 & 1.11 \\
	
$[$\AA$]$& $^{+0.27}_{-0.33}$ & $^{+0.57}_{-0.63}$ & $^{+0.69}_{-0.73}$ & $^{+1.04}_{-0.73}$ & $^{+0.24}_{-0.43}$ & $^{+0.49}_{-0.61}$ & $^{+0.55}_{-0.76}$ & $^{+0.96}_{-0.91}$ & $^{+0.64}_{-0.42}$ \\
	
&(2.67) & (2.10) & (1.90) & (1.64) & (2.65) & (2.34) & (2.08) & (1.53) & (1.21) \\

&&&&&&&&& \\	
$<$Fe$>$ & 2.57 & 2.03 & 1.81 & 1.50 & 2.61 & 2.31 & 2.04 & 2.01 & 1.16 \\
	
$[$\AA$]$&$^{+0.26}_{-0.26}$ & $^{+0.53}_{-0.54}$ & $^{+0.62}_{-0.58}$ & $^{+0.75}_{-0.61}$ & $^{+0.27}_{-0.40}$ & $^{+0.44}_{-0.61}$ & $^{+0.60}_{-0.67}$ & $^{+0.56}_{-0.70}$ & $^{+0.56}_{-0.35}$ \\
	
& (2.56) & (2.01) & (1.86) & (1.64) & (2.55) & (2.24) & (2.01) & (1.73) & (1.24) \\

&&&&&&&&& \\	
$[$MgFe$]$' & 3.12 & 2.20 & 1.82 & 1.27 & 3.17 & 2.61 & 2.20 & 1.25 & 1.07 \\
	
$[$\AA$]$& $^{+0.30}_{-0.48}$ & $^{+0.68}_{-0.62}$ & $^{+0.61}_{-0.54}$ & $^{+0.60}_{-0.37}$ & $^{+0.32}_{-0.65}$ & $^{+0.62}_{-0.72}$ & $^{+0.76}_{-0.80}$ & $^{+0.64}_{-0.29}$ & $^{+0.58}_{-0.40}$ \\
	
& (3.04) & (2.21) & (1.85) & (1.37) & (3.03) & (2.56) & (2.18) & (1.32) & (1.16) \\

&&&&&&&&&\\	
Mgb & 3.66 & 2.37 & 1.85 & 1.19 & 3.69 & 2.88 & 2.31 & 0.99 & 1.01 \\
	
$[$\AA$]$& $^{+0.51}_{-0.73}$ & $^{+0.90}_{-0.76}$ & $^{+0.83}_{-0.70}$ & $^{+0.93}_{-0.49}$ & $^{+0.60}_{-0.93}$ & $^{+0.95}_{-0.92}$ & $^{+1.07}_{-1.01}$ & $^{+0.62}_{-0.23}$ & $^{+0.78}_{-0.41}$ \\
	
&(3.55) & (2.42) & (1.91) & (1.35) & (3.55) & (2.89) & (2.33) & (1.19) & (1.19) \\ 

&&&&&&&&& \\
log($M_{BH}$/$M_{\odot}$) & 6.89 & 6.63 & 6.00 & 0.00 & 6.74 & 6.69 & 6.20 & 0.00 & 0.00 \\

(AGN) & $^{+0.65}_{-0.61}$ & $^{+0.59}_{-0.71}$ & $^{+0.76}_{-0.97}$ & $^{+0.00}_{-0.00}$ & $^{+0.52}_{-0.86}$ & $^{+0.58}_{-0.59}$ & $^{+0.59}_{-0.33}$ & $^{+0.00}_{-0.00}$ & $^{+0.00}_{-0.00}$ \\

& (7.21) & (6.94) & (6.50) & (0.00) & (6.93) & (6.98) & (6.46) & (0.00) & (0.00) \\

&&&&&&&&& \\
log(L[$\mbox{O{\sc iii}}$]/$\sigma^4$) & -1.09 & -0.90 & -0.61 & 0.00 & -1.31 & -0.79 & -0.46 & 0.00 & 0.00 \\

& $^{+0.44}_{-0.87}$ & $^{+0.90}_{-0.91}$ & $^{+0.83}_{-1.09}$ & $^{+0.00}_{-0.00}$ & $^{+1.26}_{-0.49}$ & $^{+0.67}_{-0.78}$ & $^{+0.81}_{-1.14}$ & $^{+0.00}_{-0.00}$ & $^{+0.00}_{-0.00}$ \\
&(-0.81) & (-0.19) & (0.95) & (0.00) & (-0.43) & (-0.26) & (0.10) & (0.00) & (0.00) \\

&&&&&&&&&\\	
H$\alpha$ &1.24 & 11.20 & 13.92 & 21.97 & 1.01 & 9.10 & 17.67 & 35.55 & 39.48 \\
	
(Non-AGN)$[$\AA$]$ & $^{+5.02}_{-0.75}$ & $^{+15.69}_{-8.63}$ & $^{+15.35}_{-7.96}$ & $^{+18.96}_{-10.06}$ & $^{+10.14}_{-0.55}$ & $^{+21.47}_{-7.97}$ & $^{+18.73}_{-15.25}$ & $^{+15.11}_{-17.88}$ & $^{+43.35}_{-21.91}$ \\
	
& (5.11) & (15.64) & (17.88) & (27.83) & (5.47) & (16.42) & (21.34) & (38.81) & (53.79) \\
\enddata
\label{tab:tablefive}
\end{deluxetable*}

\begin{deluxetable*}{cccccc}
\tablecolumns{12}
\tabletypesize{\scriptsize}
\tablecaption{Morphology catalogue including spectroscopic and photometric properties }
\tablewidth{0pt}
\tablehead{
\colhead{SDSS ObjID} &
\colhead{Morphology\tablenotemark{a}}	&
\colhead{R.A. [deg] \tablenotemark{b}} &
\colhead{Decl. [deg] \tablenotemark{b}} &
\colhead{Redshift} &
\colhead{$(g-r)_{0}$} 
}
\startdata
587741708323193271  &    16   &    121.37498  &     12.369875  &   0.036951199   &   0.802  \\
587727180072288271  &    22   &    28.201761  &    -8.7224340  &    0.036953799  &    0.701  \\
587726100416495842  &    22   &    229.10187  &     2.8465180  &   0.036954898   &  0.562      \\ 
587741422176698466  &    12   &    134.52182  &     22.776091  &   0.036955900   &   0.824     \\
587730022257852453  &    13   &    226.68094  &     5.3271850  &   0.036956899   &   0.909   \\ 
. &    .   &    .  &     .  &   .   &  .     \\ 
. &    .   &    .  &     .  &   .   &  .     \\
. &    .   &    .  &     .  &   .   &  .   \\ [6pt]
\hline
\hline 
$Mr_{0}$ & $\rm IsoA_r$ [arcsec] & $\rm IsoB_r$ [arcsec] & $C_{r}$ & \fracDeV  & log($M_{\ast}$/$M_{\odot}$) \\ [2pt]
\hline
-21.078   &    49.253    &   22.827    &  3.096 &      0.930     &  10.873   \\
-21.569 &      61.106  &     34.870   &  2.581 &      0.901      & 10.959   \\
-20.122   &    42.913     &  20.939      &  2.075  &       0.000   & 10.228   \\
-21.860  &     72.966    &   55.174     &  3.413  &       1.000  & 11.210   \\
-21.676  &       57.899  &       42.418 &  3.154 &      0.929     & 11.229    \\ 
. &    .   &    .  &     .  &   .   &  .     \\ 
. &    .   &    .  &     .  &   .   &  .     \\
. &    .   &    .  &     .  &   .   &  .   \\ [6pt]
\hline
\hline
$\sigma_{eff}$ $[$\kms$]$ \tablenotemark{c} & \Hb\ $[$\AA$]$ & Fe5270 $[$\AA$]$ & $<$Fe$>$ $[$\AA$]$ & Mgb	$[$\AA$]$ & $[$MgFe$]$' $[$\AA$]$ \\ [2pt]
\hline
173.860  &     1.623  &     2.674  &     2.713   &    4.368   &    3.432   \\ 
 100.353  &     3.181  &     1.806  &     1.894   &    1.879   &    1.867     \\
23.595  &     2.655  &     1.728  &     1.555   &    1.558   &    1.594     \\
 222.164  &     1.685  &     2.934  &     2.814   &    4.431  &    3.564      \\
 234.506  &     1.653  &     2.741  &     2.623   &    4.773   &    3.573     \\
 . &    .   &    .  &     .  &   .   &  .     \\ 
. &    .   &    .  &     .  &   .   &  .     \\
. &    .   &    .  &     .  &   .   &  .   \\ [6pt]
\hline
\hline
BPT class \tablenotemark{d} & log($M_{BH}$/$M_{\odot}$) \tablenotemark{e}  & \Ha\ $[$\AA$]$ \tablenotemark{f} & log(L[$\mbox{O{\sc iii}}$]/$\sigma^4$) \tablenotemark{g} & $N_{density}$ &  \\ [2pt]
\hline
 5   &    7.847  &  0.524  &    -9999  & 6 &   \\
 3   &    6.575  &  -9999   &   0.190  & 10 &   \\
 1   &    3.698  &  23.485   &   -9999   & 16 &    \\
 0   &    8.277  &  0.195  &    -9999  & 11 &  \\
 5   &    8.396  &  0.137  &    -9999  & 18 &   \\
 . &    .   &    .  &     .  &   .   &       \\ 
. &    .   &    .  &     .  &   .   &       \\
. &    .   &    .  &     .  &   .   &     \\ 

\enddata

\label{tab:tablesix}
\tablecomments{The full catalogue contains 10,233 objects. The catalogue is available in its entirety in a machine-readable form in the online journal.}
\tablenotetext{a}{E0--E5 : 10--15, S0 : 16, Sa--Sd : 21--24, SBa--SBd : 31--34, Irr : 40, unknown : 50}
\tablenotetext{b}{J2000}
\tablenotetext{c}{`-9999' is assigned when the criteria ($C_{\rm r}$ $<$ 3.5 \&\ 10 $<$ $\sigma_{eff}$ $<$ 400 $[$\kms$]$) does not satisfied.}
\tablenotetext{d}{0 : Weak-emission ($A/N \leq 1$), 1 : Star-forming, 2 : Transition region, 3 : Seyfert, 4 : LINER, 5 : Unclear ($1 < A/N < 3$)}
\tablenotetext{e}{`-9999' is assigned when $\sigma_{eff}$ is `-9999'.}
\tablenotetext{f}{`-9999' is assigned for AGN host galaxies.}
\tablenotetext{g}{`-9999' is assigned for non-AGN galaxies or log($M_{BH}$/$M_{\odot}$) is `-9999'.}
\end{deluxetable*}

\clearpage

\acknowledgments

 We are grateful to the anonymous referee for a number of clarifications that improved the quality of the manuscript. We thank Seulhee Oh for her comments to our draft. SKY acknowledges the support from the National Research Foundation of Korea to the Center for Galaxy Evolution Research (No. 2010-0027910), Doyak grant (No. 20090078756) and DRC grant of Korea Research Council of Fundamental Science and Technology (FY 2012). This project made use of the SDSS optical data.

\clearpage
\end{document}